\documentclass[journal]{IEEEtran}
\ifCLASSINFOpdf
  \usepackage[pdftex]{graphicx}
\else
  \usepackage[dvips]{graphicx}
\fi
\usepackage{amsmath}
\usepackage{amssymb}
\usepackage{subfigure}
\usepackage{color}

\begin{document}
%
\title{A Two-element Parasitic Antenna Approaching the Minimum Q-factor at a Given Directivity}
%
%
%

\author{Fabien~Ferrero,
        Leonardo~Lizzi,
        ~B.L.G.~Jonsson, and Lei Wang,
\thanks{F. Ferrero and L. Lizzi are with the University Nice - Sophia Antipolis, CNRS, LEAT, Sophia Antipolis, France e-mail: fabien.ferrero@unice.fr.}
\thanks{B.L.G. Jonsson and L. Wang are with KTH Royal Institute of Technology, at the School of Electrical Engineering, Stockholm, Sweden}
\thanks{Manuscript received \today.}}

%
%

\markboth{FERRERO+ETAL}%
{Shell \MakeLowercase{\textit{et al.}}: Bare Demo of IEEEtran.cls for IEEE Journals}
%



\maketitle

\begin{abstract}
In this paper we investigate a super-directive antenna based on a parasitic structure with a circumscribing sphere of diameter 69 mm corresponding to 0.2$\lambda$@880 MHz. The antenna is modeled, simulated, measured, and it is also evaluated against the new Q-factor bound for small antennas at a given total directivity. A maximum directivity of 7.2 dBi is measured with a radiation efficiency of -6.5 dB at 876 MHz. An intermediate directivity of 6.2 dBi is observed at 880 MHz with 20 dB front-to-back ratio and -7 dB radiation efficiency. The antenna performs well with respect to the developed fundamental bound. The results is promising for applications that require miniaturization and spatial filtering. The above antenna properties are robust and we show with measurements that the antenna preform well also when it is integrated as an autonomous unit. 

\end{abstract}


%
\IEEEpeerreviewmaketitle
\newcommand{\set}[1]{\mathrm{#1}}
\providecommand*{\mrm}[1]{\mathrm{#1}}

\newcommand{\hr}{\hat{\boldsymbol{r}}}
\newcommand{\he}{\hat{\boldsymbol{e}}}
\newcommand{\minimize}{\mathop{\text{minimize\ }}}
\newcommand{\maximize}{\mathop{\text{maximize\ }}}
\newcommand{\subjectto}{\mathop{\text{subject\ to\ }}}

\newcommand{\We}{W_{\mrm{e}}}
\newcommand{\Wem}{W_{\mrm{em}}}
\newcommand{\Wm}{W_{\mrm{m}}}

\newcommand{\RE}{\mathop{\set{Re}}}
\newcommand{\IM}{\mathop{\set{Im}}}
\newcommand{\ju}{\mathrm{j}}
\newcommand{\diff}{\mathop{\mathrm{\mathstrut{d}}}\!}
\newcommand{\vI}{\boldsymbol{I}}
\newcommand{\vR}{\boldsymbol{R}}
\newcommand{\Xe}{\boldsymbol{X}_\mrm{e}}
\newcommand{\Xm}{\boldsymbol{X}_\mrm{m}}
\newcommand{\vF}{\boldsymbol{F}}
\newcommand{\vpsi}{\boldsymbol{\psi}}
\newcommand{\vr}{\boldsymbol{r}}

\newcommand{\hx}{\hat{\boldsymbol{x}}}
\newcommand{\hy}{\hat{\boldsymbol{y}}}
\newcommand{\hz}{\hat{\boldsymbol{z}}}

\newcommand{\rP}{P_\mrm{rad}}
\newcommand{\aP}{P_\mrm{abs}}
\newcommand{\RR}{\mathbb{R}}

\newcommand{\dotp}[2]{\langle #1,#2\rangle}
\newcommand{\vJ}{\boldsymbol{J}}
\newcommand{\Le}{\mathcal{L}_\mrm{e}}
\newcommand{\Lm}{\mathcal{L}_\mrm{m}}
\newcommand{\Lem}{\mathcal{L}_\mrm{em}}

\newcommand{\lexp}[1]{\mathrm{e}^{#1}}
\newcommand{\rH}{\mathrm{H}}

\section{Introduction}

\IEEEPARstart{T}{HE EMPLOYMENT} of pattern directive antennas can be of great advantage in many communication applications. However, when the size is also a constraint, a trade off has to be found between the impedance bandwidth and the desired far-field radiation pattern. Antennas that have a spatial separation in reception i.e. a spatial filter type antenna, can mitigate jamming impacts from other nearby radiators and it can also reduce adverse multipath effects \cite{Ferrero2015,Huy2015,Patt_reconf}. 

The miniaturization of the antennas tends to limit their radiation pattern to a low-directivity regime to preserve the best bandwidth. A relationship between antenna size and maximum {\it normal} directivity and gain has been discussed in several works~\cite{Harrington,Geyi2003,Hansen2006}. In 1958, Harrington proposed a normal directivity limit: $D_H=N^2+2N$, $N=ka$ for an antenna enclosed within a sphere of radius $a$, where $k$ is the wavenumber. This limitation is based on allowing spherical vector modes of order $n\leq N$, arguing that spherical modes for $n>ka$ cannot be efficiently used in a large sphere, e.g. for $ka>1$. Super-directivity is often defined as a directivity value higher than the Harrington directivity limit. However, for small antennas for which $ka<1$, i.e. outside its validity range, it is well known that Harrington's limit underestimates the available normal directivity. As an example, note that a Huygens source has a directivity of about 4.77 dBi. To overcome this issue, both \cite{Geyi2003,Maci_Kildal} have proposed other bounds that can be used as the boundaries between normal and super directivity for $ka\geq 0$.

A review of realized antennas with respect to the size-directivity relation is given by Pigeon in \cite{pigeon}. Most of the measured superdirective antennas use an unbalanced structure with a large ground plane \cite{best,Donnel+Yaghjian,Kim+etal2012} which make them less practical for miniature applications. A promising differential structure is based on Egyptian axe-dipole with $ka=0.43$ and 5.4 dBi in peak directivity, where a prototype was presented in \cite{Tang}. Clemente et.al., \cite{clemente} presented recently a 11dBi directivity balanced parasitic antenna integrating a lumped balun component, it has $ka=1.4$. In this study, we investigate the design of a miniature balanced parasitic antenna with $ka<0.7$. 

 The development of highly directive and miniaturized antennas has been limited by challenges in the fabrication and measurements. Indeed, the input impedance and radiation pattern measurement of a miniature antenna is very challenging. Several papers have highlighted the effect of the cable on the radiation measurement \cite{massey,Haskou}.
In this work, a miniature balun is directly integrated in the radiating structure. The effect of connector and cable is studied and measurements with and without cable are presented.

It is well known that directivity is theoretically unbounded, see e.g.~\cite{Oseen1922,Uzkov1946,Riblet1948}. A normal directivity limit as given in \cite{Harrington,Geyi2003,Maci_Kildal} aims to predict where the super-directivity region starts for a given size, i.e. where the desired higher directivity start to become difficult to realize. 
In this paper we aim to explicitly quantify this `difficult to realize a certain directivity' in terms of the best realizable fractional bandwidth for each directivity. That superdirectivity is associated with small bandwidth is known, however this associated cost is often not specified for several realized antennas see e.g. the references in~\cite{pigeon}. The tools tested and developed here illustrates a method of how to, at each desired directivity, determine the best possible fractional bandwidth for all antennas that fit into a given small shape. The tools are based on stored energy bounds. 

With the introduction of the fundamental limitations on antennas~\cite{Gustafsson+etal2007} and the bounds based on stored energy \cite{Vandenbosch2010,Gustafsson+etal2012a,Gustafsson+Nordebo2013,Jonsson+Gustafsson2015,Jonsson+Gustafsson2016}, new tools have arrived to predict the Q-factor, $Q$, under different circumstances~\cite{Gustafsson+Nordebo2013,Gustafsson+etal2016a}. These bounds are determined through convex optimization~\cite{Boyd+Vandenberghe2004,Grant+Boyd2014} which have been very successful in predicting $Q$ and $G/Q$ for a range of small antennas~\cite{Gustafsson+etal2012a,Gustafsson+etal2016a,Capek+etal2016b,Jelinek+Capek2017}. Constraints on partial gain over Q, $G/Q$, for a given {\it partial} gain were introduced as a convex problem and tested in~\cite{Gustafsson+Nordebo2013,Gustafsson+etal2016a}. Most of the investigated antennas are antennas that can be printed/design in a single layer, or a layer above a grounded plane~\cite{Tayli+Gustafsson2015}. We note that all these latter bounds are valid for arbitrary shapes of the antenna. An antenna has electric and magnetic static polarizabilities see e.g.~\cite{Schiffer+etal1949,Sjoberg2009b}, which appears in the small antenna limit as a fundamental constant in the $G/Q$-bounds and $Q$-bounds \cite{Gustafsson+etal2007,Stuart+Yaghjian2010,Jonsson+Gustafsson2015}. In the present paper we develop bounds on the Q-factor for a given total directivity in a given direction. We also include comparisons between the different methods to bound the Q-factor both for the total and the partial directivity. 

In antenna design it is often the gain of the antenna that is a desired design parameter. Furthermore, simulations and measurements determine in general the bandwidth, not the Q-factor for the design. For loss-less antennas it is well known that partial gain and partial directivity coincide. However realized antennas are seldom loss-less, but well designed antennas tend to have low losses. The relation between relative (or fractional) bandwidth to Q-factors and vice-verse is discussed in e.g.~\cite{Yaghjian+Best2005,Gustafsson+Jonsson2015a}. It is known that $Q^{-1}$ and relative bandwidth correlate, but it is not quite predictive~\cite{Gustafsson+Jonsson2015a,Gustafsson+Nordebo2006}. However, electrically small antennas tend to have a better correlation between $Q^{-1}$ and the relative bandwidth. 

One of the goals of this paper is to present a new miniaturized parasitic element antenna and its  measurements. The proposed antenna design is composed with two planar structures. The two planar structures have the same geometry which open the possibility to use this solution in a reconfigurable antenna. The antenna design integrates a balun to remove undesired currents on the shielding of the coaxial cable feeding the antenna. 
Another aim is to use the fundamental bounds adapted for the designed antenna to determine how it matches with the bandwidth predictions from stored energies at a given directivity. Applying the design development and comparisons with the bounds, we fabricate a prototype which is measured and validated. 

The paper is outlined as follows: Section II presents the geometry and antenna design. Section III, introduce the stored energy approach and determines lower bounds for directivity as a function of the Q-factor for the given geometry. Section IV presents a practical realization and associated measurements. The paper ends with a discussion and conclusions.

\section{ANTENNA DESIGN}

\subsection{Proposed structure}

\begin{figure}[t]
\centering{\includegraphics[width=75mm]{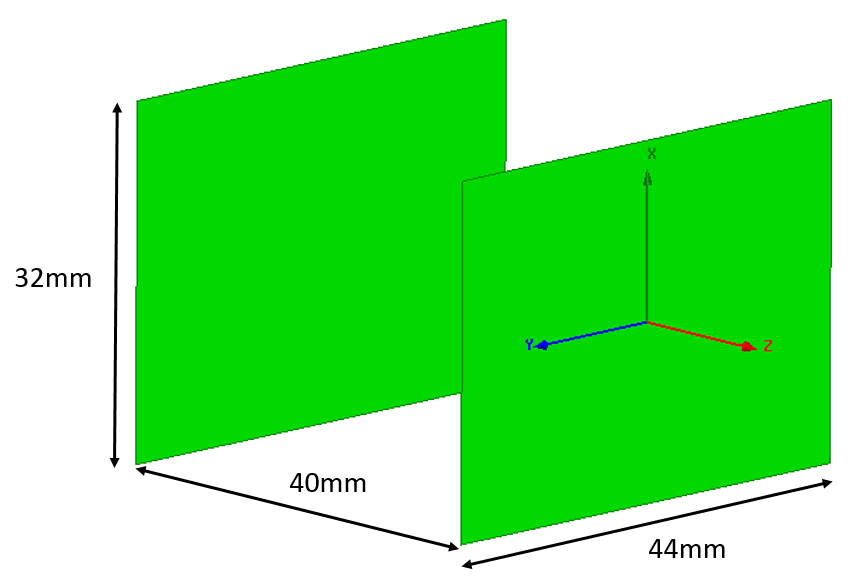}}
\caption{Geometry 1 of the parasitic element structure.}
\label{paras}
\end{figure}
In order to push down the limit while keeping an acceptable efficiency,  a $ka < 0.7$ is targeted, meaning that the antenna has to be circumscribed in a sphere of 69 mm diameter. The proposed structure is presented in Fig. \ref{paras}, the rectangular plate has a width of 32 mm and a length of 44 mm and the distance between the two plates is 40 mm.

We focus our structure on a parasitic array with a single feed element combined with one parasitic element. The reason for this is that separately fed elements require accurate control of the amplitude and phase on the element input excitations. Such super-directive antenna array require a complicated feeding network which will add complexity for a practical realization.
In order to simplify the prototyping, the structure is based on two planar rectangular face.
The dimensions of the rectangular plate are inspired by the work of Gustafsson \cite{Gustafsson2009} who shows that a ratio between the length $l_1$ and the width $l_2$ of $1.4$ give optimal performance for the $D/Q$ criteria.
Moreover, in order to maximize the occupancy of the sphere around the antenna, a distance between the plates of ($l_1$+$l_2$)/2 is chosen.

\subsection{Single element Antenna design}

\begin{figure}[t]
\centering{\includegraphics[width=75mm]{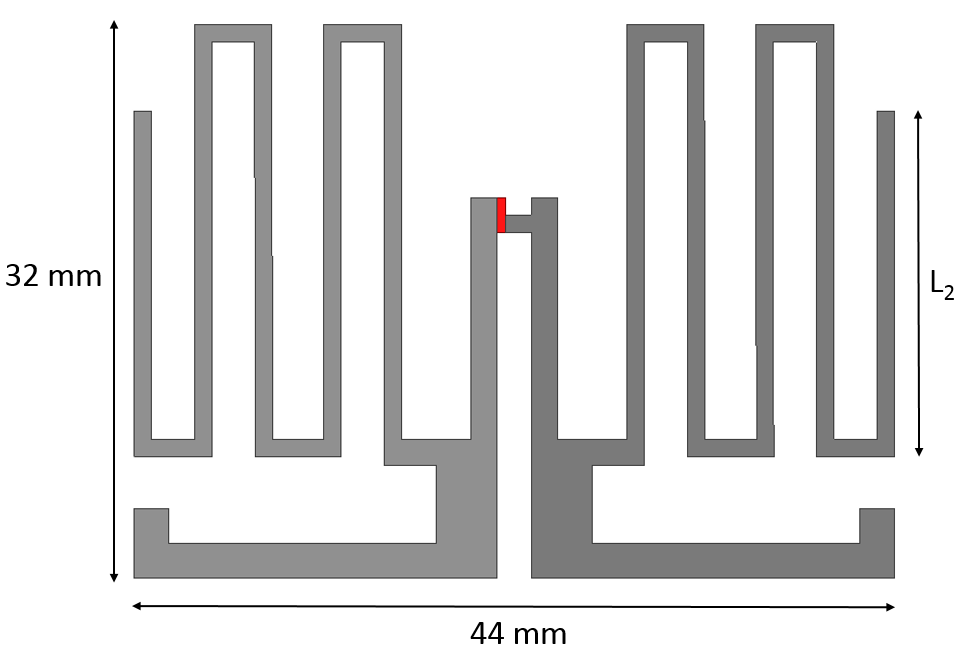}}
\caption{Geometry of the single element structure.}
\label{single1}
\end{figure}

\begin{figure}[th]
\centering{\includegraphics[width=75mm]{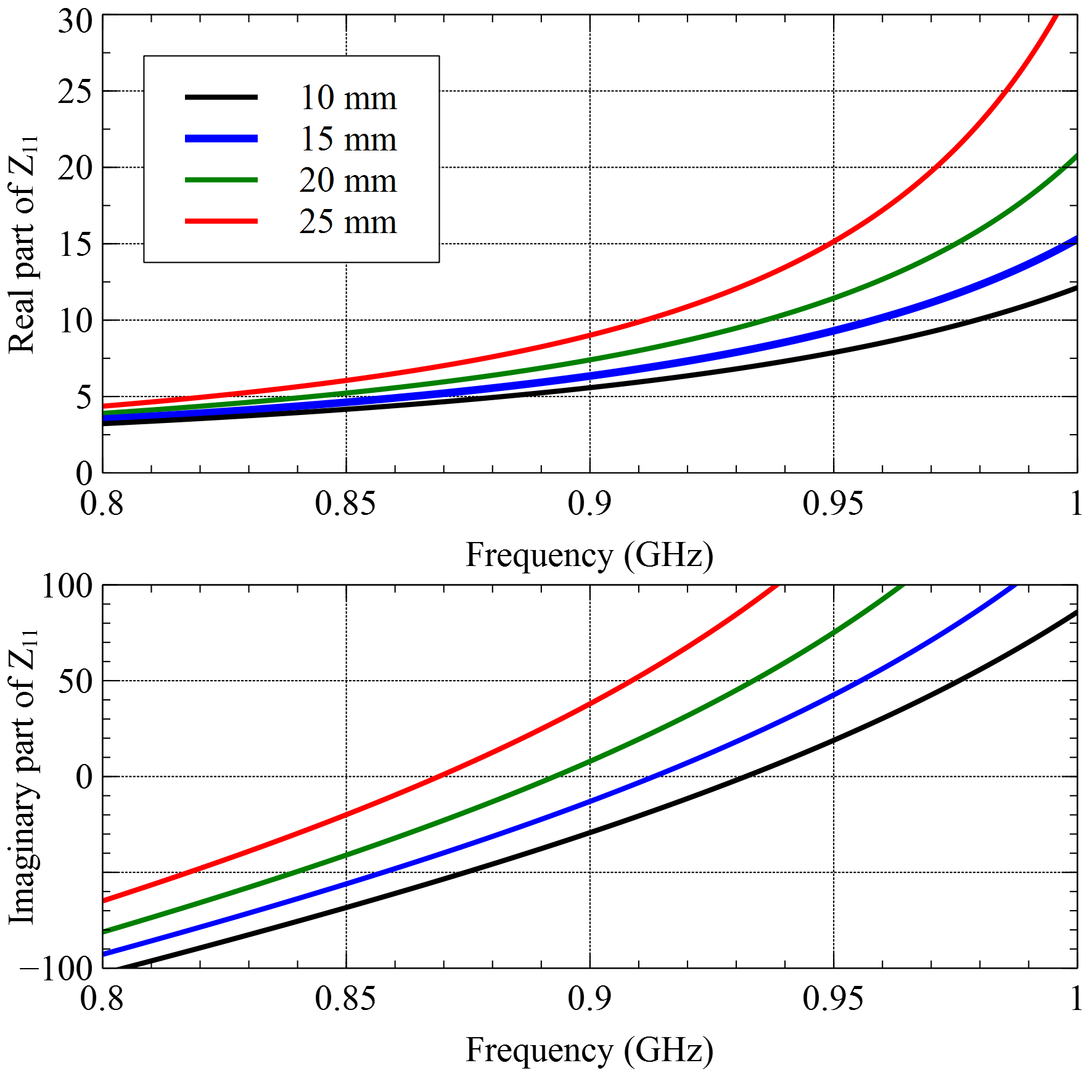}}
\caption{A simulation of the real and imaginary part of the antenna input impedance for a  single element versus frequency, for different length of the meander line, the $L_2$-length in see Fig.~\ref{single1}.}
\label{Z11}
\end{figure}

Our approach is based on the design of a miniature electric dipole antenna. In order to miniaturize the dipole element, the two branches are strongly meandered as shown in Fig. \ref{single1}. As it can be seen, the structure is symmetrical with respect to the vertical axis. A balun, constituted by an open slot with open stub, is integrated within the dipole to enable measurements with a coaxial cable. The feeding point is marked in red.  
The resonant frequency of this structure can be tuned by changing the length of the meandered dipole as shown in Fig. \ref{Z11}. In order to tune the single antenna resonance at 0.92 GHz, we chose $L_2=20$ mm (see Fig.~\ref{single1}). As shown in Fig.~\ref{S11}, the single element is not intended to be matched to 50$\Omega$ when used alone.

\subsection{Two elements parasitic Antenna design}

On the second surface of the geometry shown on Fig.~\ref{paras}, a reflective element is designed. The parasitic element has exactly the same geometry as the driven element, with the feeding port short-circuited. This approach is interesting for potential reconfigurability by switching the feed and the short circuit load.

\begin{figure}[t]
\centering{\includegraphics[width=75mm]{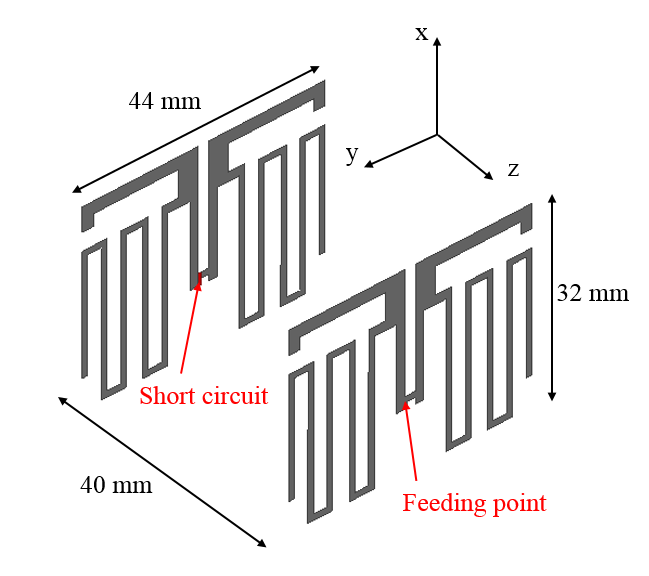}}
\caption{The geometry of the two elements parasitic structure.}
\label{double}
\end{figure}

Having such a closely spaced electric resonator imply a strong mutual coupling between the elements, which can be analyzed from the self and mutual impedance. The quantities $Z_{11}$, $Z_{22}$ are the self impedances of the two antennas, equivalent to the input impedances of the isolated antennas, clearly $Z_{11}=Z_{22}$. $Z_{12}$ and $Z_{21}$ are the mutual impedances: a current in one dipole will induce a voltage in the second one and reciprocity implies that $Z_{12}$ = $Z_{21}$. 
\begin{equation}
V_{1} = Z_{11}I_1 + Z_{12}I_2, \ \
V_{2} = Z_{21}I_1 + Z_{22}I_2,
\end{equation}

If the first antenna element is driven and the second is parasitic (short-circuited), then $V_2=0$ and equation below can be extracted:
\begin{equation}
Z_{in}=\frac{V_1}{I_1}=Z_{11}\left(1 - (\frac{Z_{21}} {Z_{22}})^2\right), \label{eq:Zin}
\end{equation}

\begin{figure}[t]
\centering{\includegraphics[width=75mm]{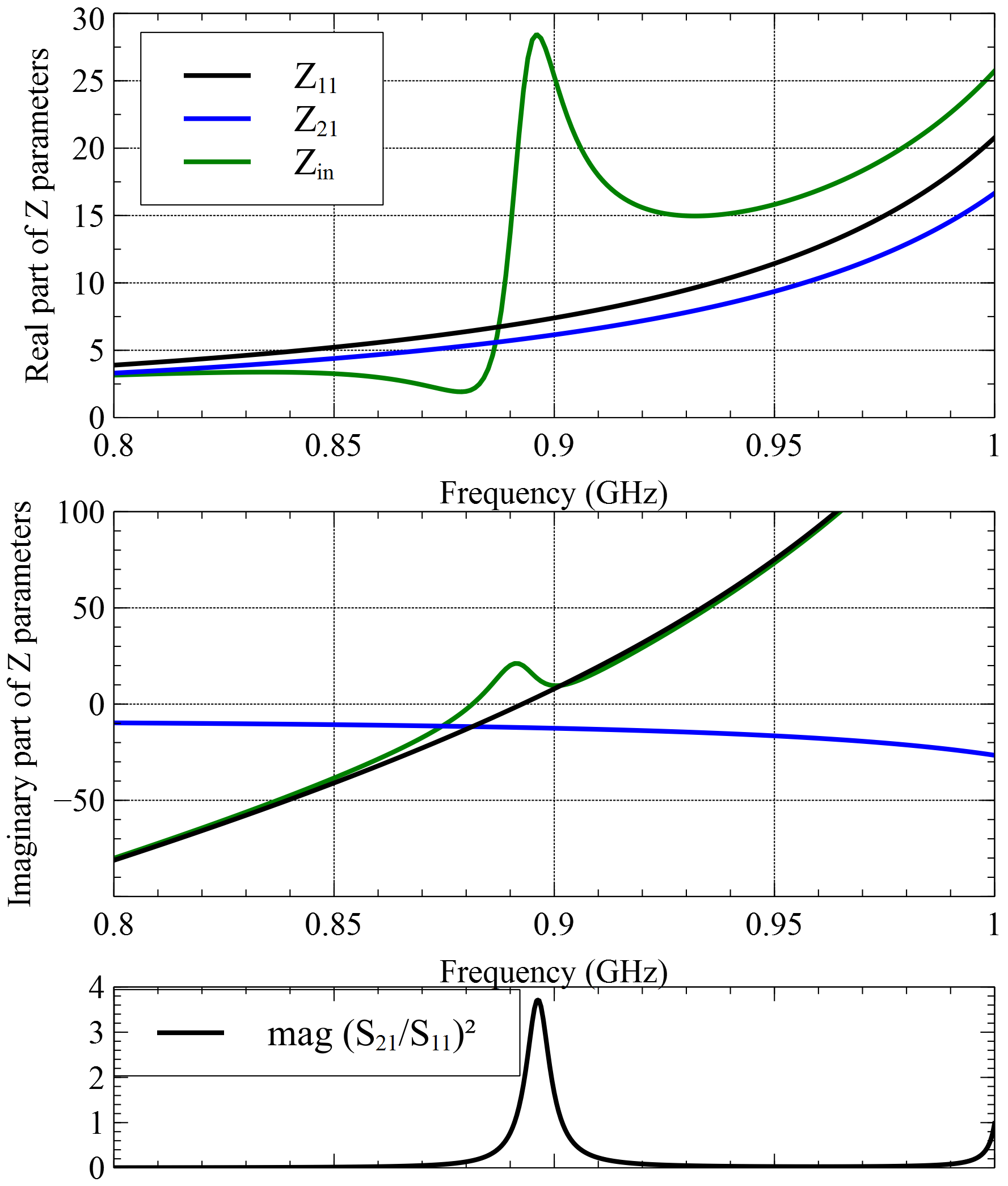}}
\caption{Impedance and scattering parameters for the parasitic geometry given in Fig.~\ref{double}.}
\label{Z21}
\end{figure}

The proposed structure is modeled with HFSS EM simulator and impedance parameters are extracted and presented in Fig.~\ref{Z21}. The ratio $Z_{21}/Z_{22}$ quantifies the effect of the coupling and the deviation of $Z_{in}$ from $Z_{11}$. It can be seen that this ratio is maximal around the antenna resonance.

\begin{figure}[t]
\centering{\includegraphics[width=75mm]{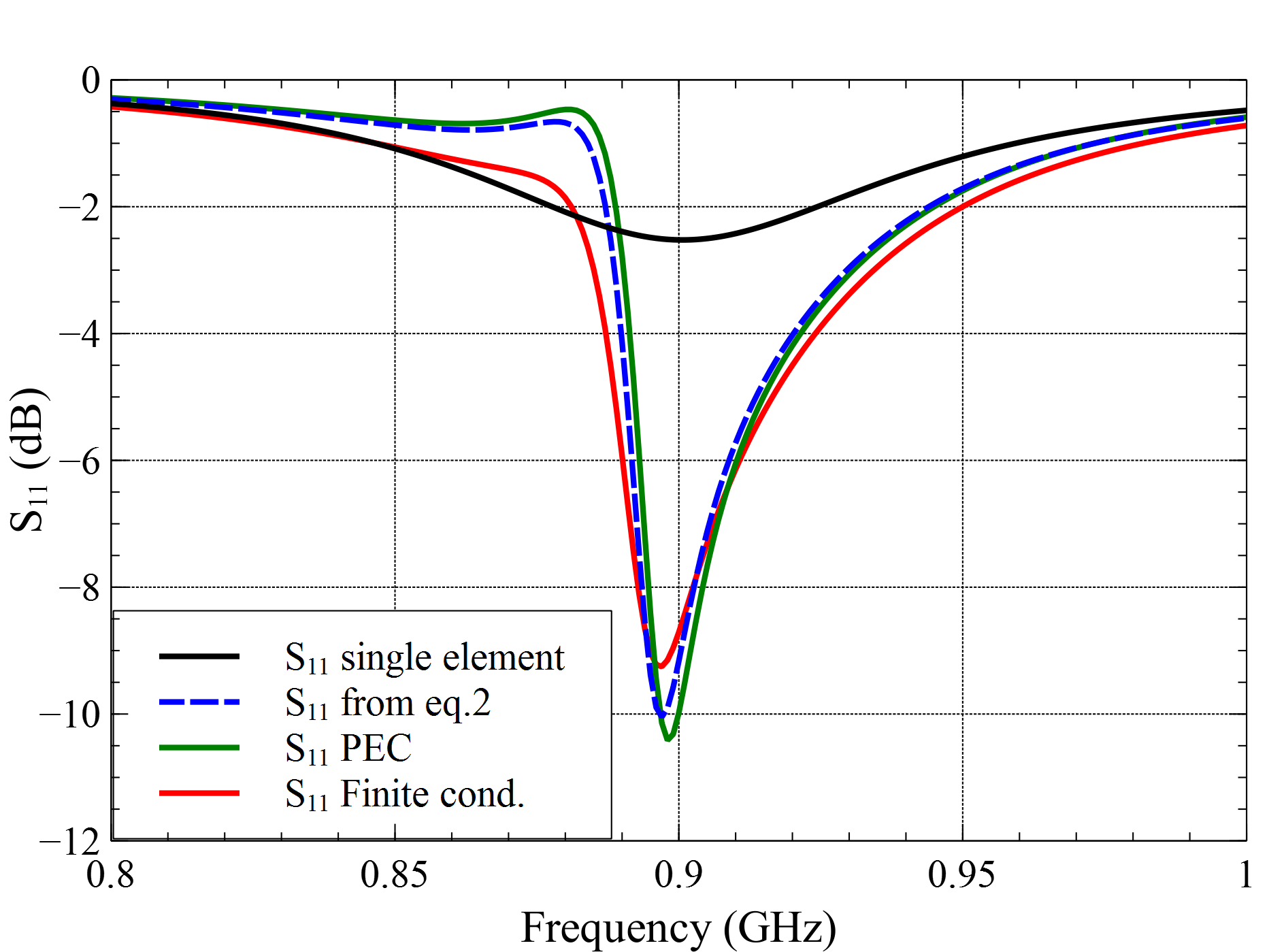}}
\caption{The antenna reflection coefficient, with respect to a $Z_0=50\Omega$-feed.}
\label{S11}
\end{figure}

The reflection coefficient of the two element antenna with a driven element and a short-circuited elements is computed with perfect electric conductor (PEC). This result is compared with Eqn.~\eqref{eq:Zin} in Fig.~\ref{S11} and a very good agreement is observed. A finite conductivity results in a similar matching level and resonance frequency with a lower Q-factor.

\begin{figure}[t]
\centering{\includegraphics[width=80mm]{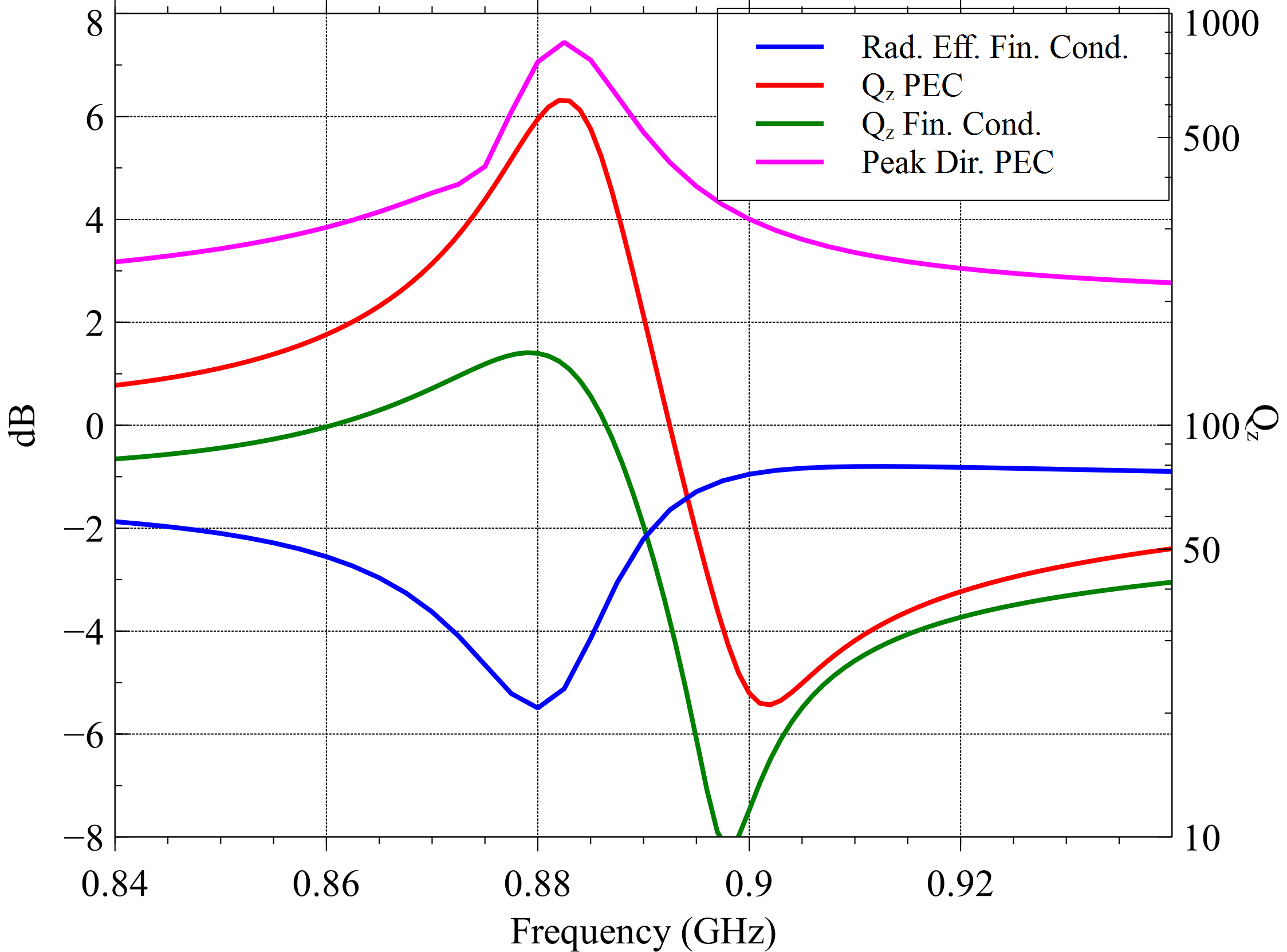}}
\caption{$Q_Z$, directivity and radiation efficiency versus frequency.}
\label{Qz_supD}
\end{figure}

In order to analyze the radiation properties of the antenna, a full 3D Electromagnetic simulation is performed with perfect electric conductor (PEC). The simulation of the maximal directivity versus the Q-factor is computed using the $Q_Z$ formula in \cite{Yaghjian+Best2005} see Fig.~\ref{Qz_supD}. The effect of conductive losses is investigated in a second simulation ($\sigma_{Cu} = 5.8\cdot 10^7$ S/m), and the $Q_Z$ and radiation efficiency is plotted in Fig.~\ref{Qz_supD}. A maximum directivity of 7 dBi is obtained, and it corresponds to the maximal $Q_Z$ for PEC and Finite conductivity model, and minimal radiation efficiency, as shown in Fig. \ref{Qz_supD}.
Thus, it is interesting to observe how the increase of directivity is correlated with the increase of quality factor and a decrease in the radiation efficiency. The variation of $Q_Z$ versus frequency for PEC is extremly sharp, indeed $Q_Z=22$ @0.9 GHz \& $Q_Z$=500 @0.88 GHz.

\begin{figure}[t]
\centering{\includegraphics[width=80mm]{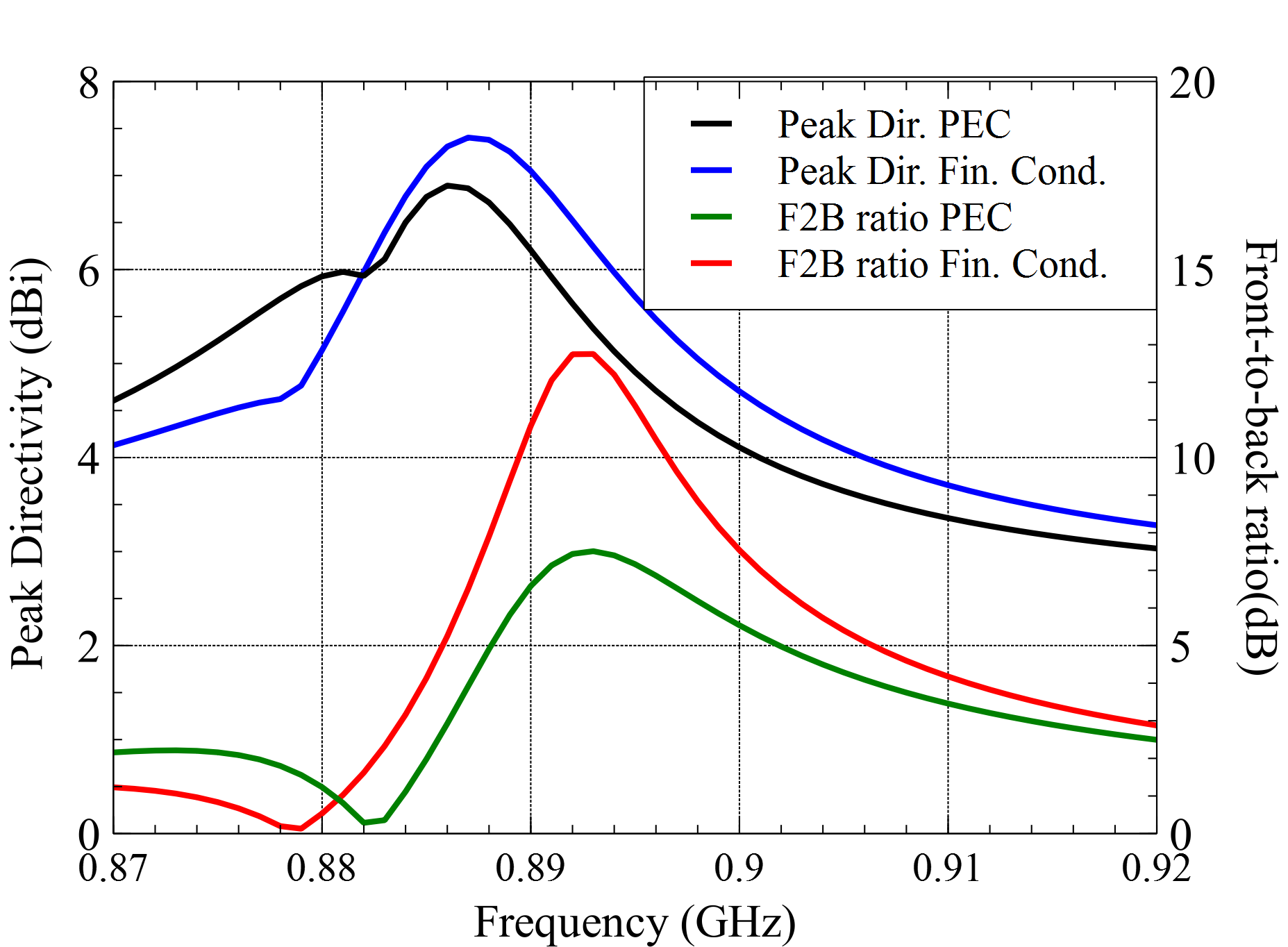}}
\caption{Simulated Directivity and Front to Back ratio versus frequency for loss-less and lossy model in Fig.~\ref{double}.}
\label{F2B}
\end{figure}

The spatial filtering capabilities of the antenna can be analyzed to the first order by using the front-to-back ratio (F2B) in the $z$-direction, as shown in Fig. \ref{F2B}. EM simulation are performed both for PEC and finite conductivity materials. It is interesting to observe that the directivity, and even more, the F2B ratio are very different between the loss-less and lossy case. Replacing PEC with the conductivity of copper results in that the ratio of the peak front-to-back ration increase by 7dB. The directivity is also increased for the models with losses as shown in \ref{F2B}.

This first antenna model, Fig.~\ref{double}, shows that a directive miniature antenna with $ka<0.7$ can be designed with directivity higher than 7dB and F2B ratio higher than 12 dB. Given such a design, how close is it to optimal ? From a single realization it is impossible to state much about optimality of absolute performance bounds. Indeed, there are an infinite variation of possible structures that can be placed on the two planar plates, which may provide better results. Thus, we compare in the next section our concept antenna with the stored energy performance bounds.

\section{Q-factor vs DIRECTIVITY limits}
\subsection{Stored energy approach}

Given the simulated and measured antenna performance as described above, it is interesting to compare these results with a theoretical approach. To enable such a comparison we will study the lossless case. The aim is to find bounds on the Q-factor with for a given total directivity. The bounds on the Q-factor are based on the electric $\We$ and magnetic $\Wm$ stored energy, a concept that has been thoroughly discussed in the literature, \cite{Collin+Rothschild1964,Fante1969,Geyi2003,Yaghjian+Best2005,Vandenbosch2010,Gustafsson+Jonsson2013,Gustafsson+Jonsson2015a,Jonsson+Gustafsson2016}, 
\begin{equation}
\We=\We^{(0)}+\Wem, \ 
\Wm=\Wm^{(0)}+\Wem,
\end{equation}
where 
\begin{align}
\We^ {(0)} &= \frac{\mu_0}{4k^ 2}\int_S\int_{S} (\nabla_1\cdot \vJ_{1})(\nabla_2\cdot\vJ_{2}^*)\frac{\cos(kR)}{4\pi R}\diff S_1\diff S_2, \\
\Wm^ {(0)} &= \frac{\mu_0}{4}\int_S\int_{S} \vJ_{1}\cdot\vJ_{2}^*\frac{\cos(kR)}{4\pi R}\diff S_1\diff S_2,
\end{align}
and
\begin{multline}
\Wem=\frac{-\mu_0}{4k}\int_S\int_S \big(k^2\vJ_{1}\cdot\vJ_{2}^ *-(\nabla_1\cdot\vJ_{1})(\nabla_2\cdot \vJ_{2}^*)\big)\cdot\\\frac{\sin(kR)}{8\pi}\diff S_1 \diff S_2.
\end{multline}
Here $\vJ_m=\vJ(\vr_m)$, $m=1,2$ is the electric surface current density over the domain and $\vr_m$ are a vectors in $\RR^3$, $k$ is the wave number, $R=|\vr_1-\vr_2|$ and $\mu_0$ is the free space permeability.  

The lower bound on the Q-factor for given antenna or antenna surrounding shape $S$ is defined by
\begin{equation}
Q=\min_{\vJ} \frac{2\omega \max(\We,\Wm)}{\rP+\aP},
\end{equation}
where $\aP$ is the Ohmic-losses in the structure and where the radiated power is defined as
\begin{multline}
\rP=\eta_0\int_S\int_S (k^2\vJ_1\cdot\vJ_2 + (\nabla_1\cdot\vJ_1)(\nabla_2\cdot\vJ_2))\cdot \\
\frac{\sin(kR)}{8\pi kR}\diff S_1\diff S_2.
\end{multline}
We want to determine the lower bounds on $Q$ under constraints on  the total directivity, $D>D_*$, for some desired $D_*$. 

The partial gain and partial directivity in the direction $\hr=\vr/|\vr|$ with polarization $\he$ are defined by~\cite{IEEE2013}:
\begin{equation}
G(\hr,\he)=\frac{4\pi P(\hr,\he)}{\rP+\aP},\ D(\hr,\he)=\frac{4\pi P(\hr,\he)}{\rP},
\end{equation}
where 
\begin{multline}
P(\hr,\he)=\frac{1}{2\eta_0}|\he^*\cdot \vF_{\mrm{E}}(\hr)|^2\\=\frac{\eta_0 k^2}{32\pi^2}|\int_S \he^*\cdot \vJ(\vr_1)\lexp{\ju k\hr\cdot\vr_1}\diff S_1|^2,
\end{multline}

We simplify the above outlined problem by considering the total directivity in a given direction $D(\hr)$, where $\hr$ is a unit radial vector for a given spherical direction, and sweep $\hr$ over a range of angles.  The total directivity $D(\hr)$ is defined by 
\begin{equation}
D(\hr)= \frac{4\pi (P(\hr,\he_1)+P(\hr,\he_2))}{\rP}.
\end{equation}

To numerically determine these quantities we use the method-of-moment approach with RWG-basis functions, and note that the desired stored energies are very similar to the EFIE-integral equation terms. An extended discussion of such an implementations is given in~\cite{Gustafsson+etal2016a}.

\subsection{The minimization problems}

We aim to find a lower bound on $Q$ for a desired total directivity $D_*$, in the lossless case, e.g. $\aP=0$. That is we would like to solve the minimization problem: 
\begin{align}\label{q}
\minimize_I &\frac{2\omega \max (\We,\Wm)}{\rP}, \\
\subjectto &\max_{\theta,\phi} \frac{4\pi (P(\hr,\he_1)+P(\hr,\he_2))}{\rP}\geq D_*,\label{q1}
\end{align}
where $\hr(\theta,\phi)$ is the radial unit direction and $\he_1(\theta,\phi)$ and $\he_2(\theta,\phi)$ are two orthogonal directions on the far-field sphere. Here $\{\hr,\he_1,\he_2\}$ form an orthogonal triplet at each direction parametrized by $(\theta,\phi)\in [0,\pi)\times[0,2\pi)$. Clearly we can simplify the problem by determining the bound for a given direction $\hr$:
\begin{align}\label{Q}
\minimize_I &\frac{2\omega \max (\We,\Wm)}{\rP}, \\
\subjectto &\frac{4\pi (P(\hr,\he_1)+P(\hr,\he_2))}{\rP}\geq D_*.
\label{Q1}
\end{align}
Solving several problems with different $\hr$-directions in \eqref{Q}-\eqref{Q1} will give an estimate of the solution to~\eqref{q}-\eqref{q1}. A consequence of the given direction in ~\eqref{Q}-\eqref{Q1} is that for a current such that $D(\hr)\geq D_*$ at minimum $Q$, it is possible that the directivity in another direction can be larger than $D(\hr)$ for the minimal Q-factor, for a discussion see~\cite{Jonsson+etal2017}. The problem~\eqref{Q}-\eqref{Q1} can be formulated as a quadratically constrained quadratic problem. Such problems can be NP-hard. 

It was shown in ~\cite{Gustafsson+Nordebo2013} that bounds on $Q$ for a given {\it partial} directivity $D(\hr,\he)\geq D_0$ i.e. 
\begin{align}\label{goq}
\maximize_I & \frac{4\pi P(\hr,\he)}{2\omega\max (\We,\Wm)}, \\
\subjectto &\frac{4\pi P(\hr,\he)}{\rP}\geq D_0,\label{goq1}
\end{align}
can be formulated as a convex problem. This feature makes it fast to solve for arbitrary antennas of arbitrary shapes. 

It is interesting to compare these different approaches to obtain a lower bound on the Q-factor, and how they differ for a given geometry. By solving both~\eqref{Q}-\eqref{Q1} and \eqref{goq}-\eqref{goq1} we can compare the effects of bounds on partial and total directivity. 

\subsection{Description of multi-element structures}

To determine a lower bound on the Q-factor for a given directivity for a given antenna geometry, note that increasing the support of the currents reduces the Q-value. This follows since we allow more possible currents on the domain. Thus to determine the bounds on the parasitic antenna structure we replace the detailed geometry as given in e.g. Fig.~\ref{double} with the structure in Fig.~\ref{paras}. The rectangles are electrically disconnected. All finite energy currents are allowed on these rectangles in the optimization process. With this geometry the minimum sphere circumscribing the antenna is equal to $\sim$0.2$\lambda_0$, corresponding to a $ka\approx 0.64$ @0.89GHz. This 3D-structure is one of the first multi component geometries that have been investigated with the Q-factor bounds see also \cite{Wakasa2016}.

\subsection{Results of the lower-bound estimates}

\begin{figure}[t]
\centering{\includegraphics[width=80mm]{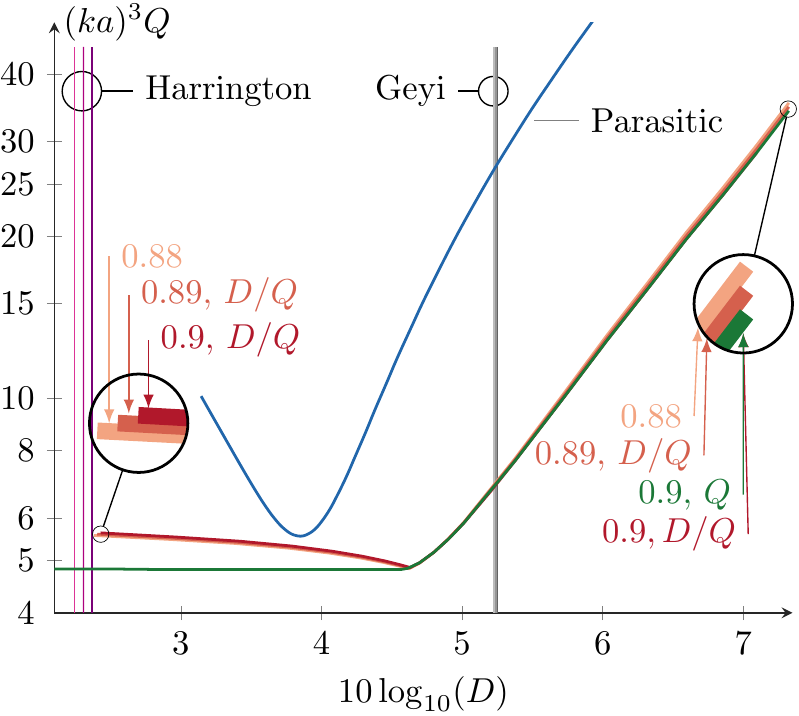}}
\caption{The blue line is a HFSS simulation of a two element loss-less parasitic antenna in Fig~\ref{double}.
The red, orange and yellow curves depict the partial  polarization-optimization~\eqref{goq}-\eqref{goq1} predicting lowest $Q$ for the given directivity, for the shape show in Fig.~\ref{paras}. The green line corresponds to the prediction of lowest $Q$ bound utilizing~\eqref{Q}-\eqref{Q1}, for the shape~Fig.~\ref{paras}. The vertical lines at $D_H=\{2.24, 2.31, 2.36\}$ dBi are Harringtons limit, and the vertical lines at $D_G=\{5.23,5.24,5.25\}$ dBi are Geyi's limit  for $f=\{0.88, 0.89, 0.9\}$ GHz for the given structure.}
\label{limits}
\end{figure}
We solve the~\eqref{goq}-\eqref{goq1} problem by utilizing cvx \cite{Grant+Boyd2014} for a range of different tested radiation directions $\hr$ as well as different minimal partial directivity $D_0$. The selected radiation directions are on a circle in the $xz$-plane with a polarization along the $y$-axis. 

For each radiation direction, we sweep the directivity over [2-7] dBi and determine the associated Q-factor. For each such sweep we extract the lowest Q-factor bound. The results are depicted in [yellow,orange,red]-colors in Fig.~\ref{limits} corresponding to the frequencies  [0.88, 0.89, 0.9] GHz.
The curves overlap due to the scaling the $Q$-factor with $(ka)^3$, utilizing that small antennas have a lower bound on the Q-factor that is proportional to $(k^3\gamma_{\text{max}})^{-1}$, where $\gamma_{\text{max}}$ is an appropriate combination of the polarizability of the structure see~\cite{Jonsson+Gustafsson2015}. The uncertainty in the depicted graphs is about 3\%. 

The initial reduction in smallest $Q$-value for an increasing partial directivity problem has its roots in that we consider a fixed polarization for different observation directions. As a difference the solution~\eqref{Q}-\eqref{Q1} allow a both polarizations that thus lowers the Q-factor at low directivites. In the Fig.~\ref{limits} we have also included a HFSS simulation of a two-element loss-less parasitic element antenna in Figure~\ref{double} in blue. This graph is calculated as a parametric curve by sweeping the frequency near its minimum at 0.895 GHz, see Fig.~\ref{S11} and using directivity and $Q_Z$ in Fig.~\ref{Qz_supD} from full EM-simulations over the desired frequencies.
A comparison between the cases with and without the conductive losses of copper is shown in Fig.~\ref{F2B} and ~\ref{limits} for radiation efficiency, directivity and Q-factor. 
It is satisfactory to observe how close to the limit is the proposed antenna concept.
Of course, the proximity with fundamental limits is achieved only for a narrow bandwidth corresponding to a directivity of 4dBi.

We also solve~\eqref{Q}-\eqref{Q1} for $f=0.9$ GHz only. The resulting lower bound on $Q$ is given in green for the frequency $0.9$ GHz. We note that above $D\approx 4.5$ dBi both curves coincide for the considered range.  Below this value on the directivity we observe essentially a straight line for the lower bound on $Q$ with respect to increase in directivity. We conclude that if we allow both polarizations, then it is possible to find a radiation direction with a fixed $Q$-value for with total directivity below $D<4.5$ dBi. It is our conclusion that the structure in Fig.~\ref{paras} has its start of a bandwidth associated costs of super-directivity at the region above $D\approx 4.5$ dBi, which is rather close to the directivity of a Huygens source, where suddenly the bandwidth rapidly decrease. This transition point is in the neighborhood of Geyi's `normal' bound on $D$ (vertical gray lines) at $D_G\approx 5.2$ dBi. A detailed discussion of methods to solve the minimization problem (14)-(15) is described in~\cite{Jonsson+etal2017}.

\begin{figure}[t]
\centering{\includegraphics[width=90mm]{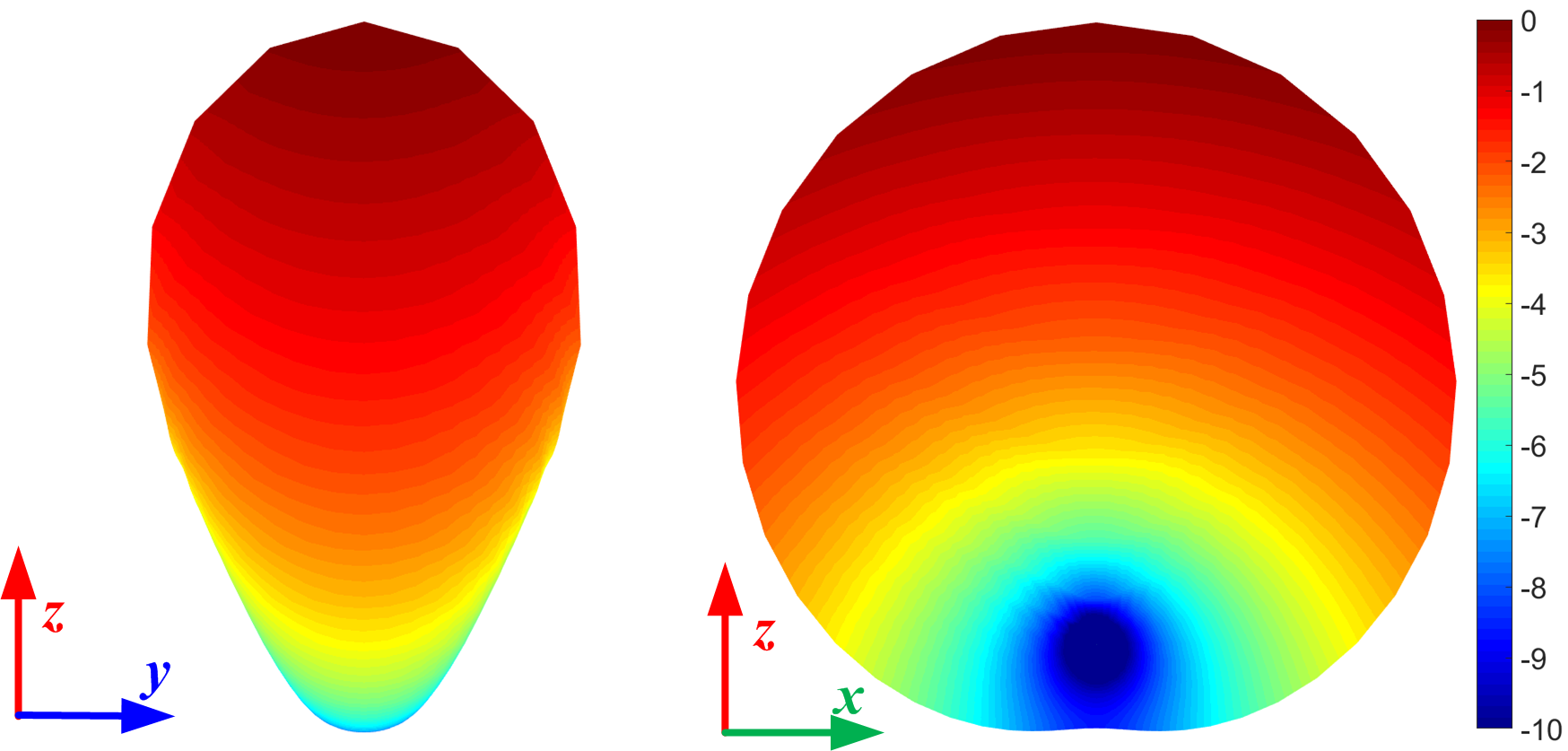}}
\caption{Radiation pattern with $\hr=\hz$ for an optimal current at 880 MHz when $D=4.8$ dBi.}
\label{aradpattern}
\end{figure}

The radiation pattern of the $\hr=\hz$ optimization of~\eqref{Q}-\eqref{Q1} at $D=4.8$ dBi are shown in Figure~\ref{aradpattern}.


\section{Antenna prototyping and measurement}

\subsection{Substrate and fabrication}

\begin{figure}[t]
\centering{\includegraphics[width=75mm]{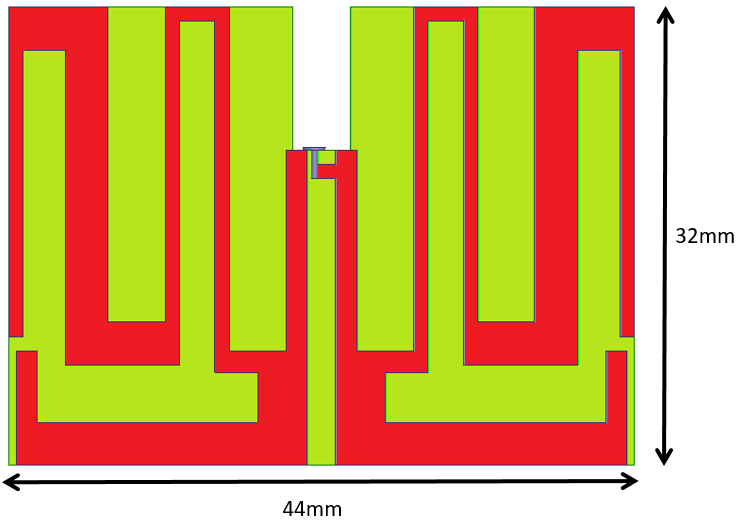}}
\caption{Geometry 2 of the single element structure taking into account substrate effect}
\label{single}
\end{figure}

\begin{figure}[t]
\centering{\includegraphics[width=92mm]{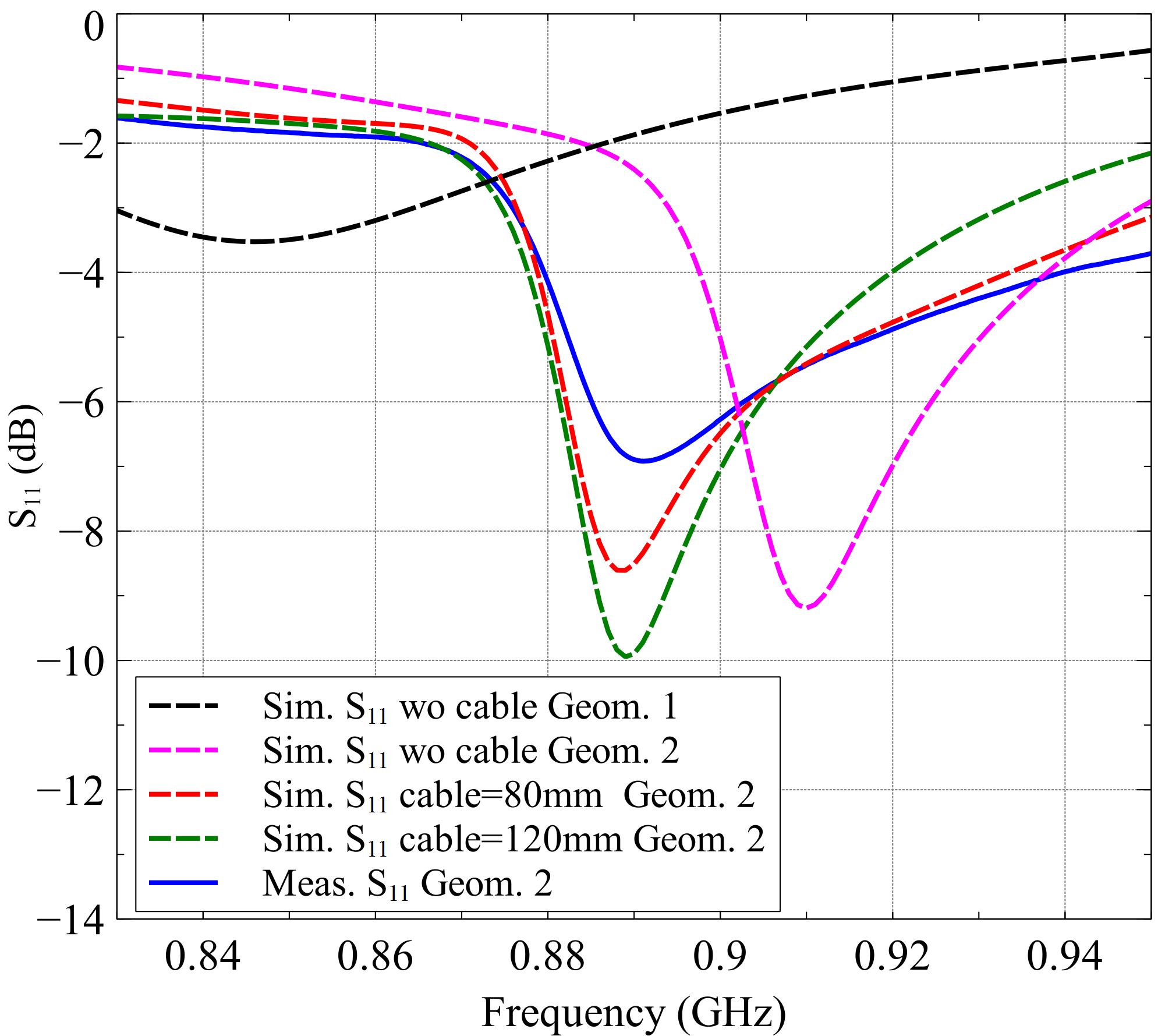}}
\caption{Simulated and measured reflection coefficient, for geometries 1 and 2 without cable and with cable of two different lengths.}
\label{s11_mesure_simu}
\end{figure}

In order to realize a prototype of this concept, a substrate is required for the fabrication. In order to limit as much as possible the effect of a such substrate and prototype cost, a 0.4mm-thick FR4 Epoxy substrate is selected. However, despite the small thickness of the substrate, the geometry Nr.~1 is strongly affected as shown in Fig. \ref{s11_mesure_simu}. The resonance frequency of the structure is shifted down to $0.845$ GHz and the input impedance becomes unmatched. An improved geometry: Nr.~2 is proposed as shown in Fig. \ref{single}. The length of the dipole branches are shortened and different widths on the branches are used to improve the reflection coefficient down to the -6dB criteria. The dipole is printed on a 0.4mm-thick FR4 Epoxy substrate with material parameters $\epsilon_r=4.4$ and $\tan\delta=0.02$.

The prototype is manufactured and connected using a thin coaxial cable. Several constraints appear with the realization of the antenna. Of course, the dielectric and conductor are both lossy, and the antenna has to be fed by a generator, connected through a cable.
The effect of the cable on the matching is presented in Fig. \ref{s11_mesure_simu}. It can be seen that despite the balun, the effect of the cable is significant, especially considering the narrowness of the frequency bandwidth. 
A first assumption in observing the cable effect as depicted in Fig.~\ref{s11_mesure_simu}, was to blame the balun for wrong operation. However, simulations with cable longer than 80mm show a very weak sensitivity from the cable length together with a good radiation pattern. We believe that this effect is due to the influence of the cable in the reactive near field of the antenna, leading to this $3$\% relative frequency decrease.
For a cable length of 80 mm and 120 mm, the resonance frequency is stable, and the HFSS simulation predicts a minimal reflection coefficient of about -9 dB at 0.89 GHz as shown in Fig. \ref{s11_mesure_simu}. A good agreement is found with the measurement.
The quality factor extracted using the $Q_Z$ formula in \cite{Yaghjian+Best2005} from impedance simulation and measurement are presented in Fig.~\ref{Dir}. A good agreement is obtained between simulated and measured data. The peak realized gain of 2 dBi is obtained at 0.885 GHz with a 5.8 dBi peak directivity and $Q_Z=35$ quality factor.

The antenna radiation characteristics are measured using two different technique: 
first with a classical coaxial cable feeding the driven element through the integrated balun;
then with an integrated transciever in order the demonstrate the capabilities of this structure to operate autonomously while keeping the radiation pattern performances.

\subsection{Measurement with coaxial cable }
\begin{figure}[t]
\centering{\includegraphics[width=85mm]{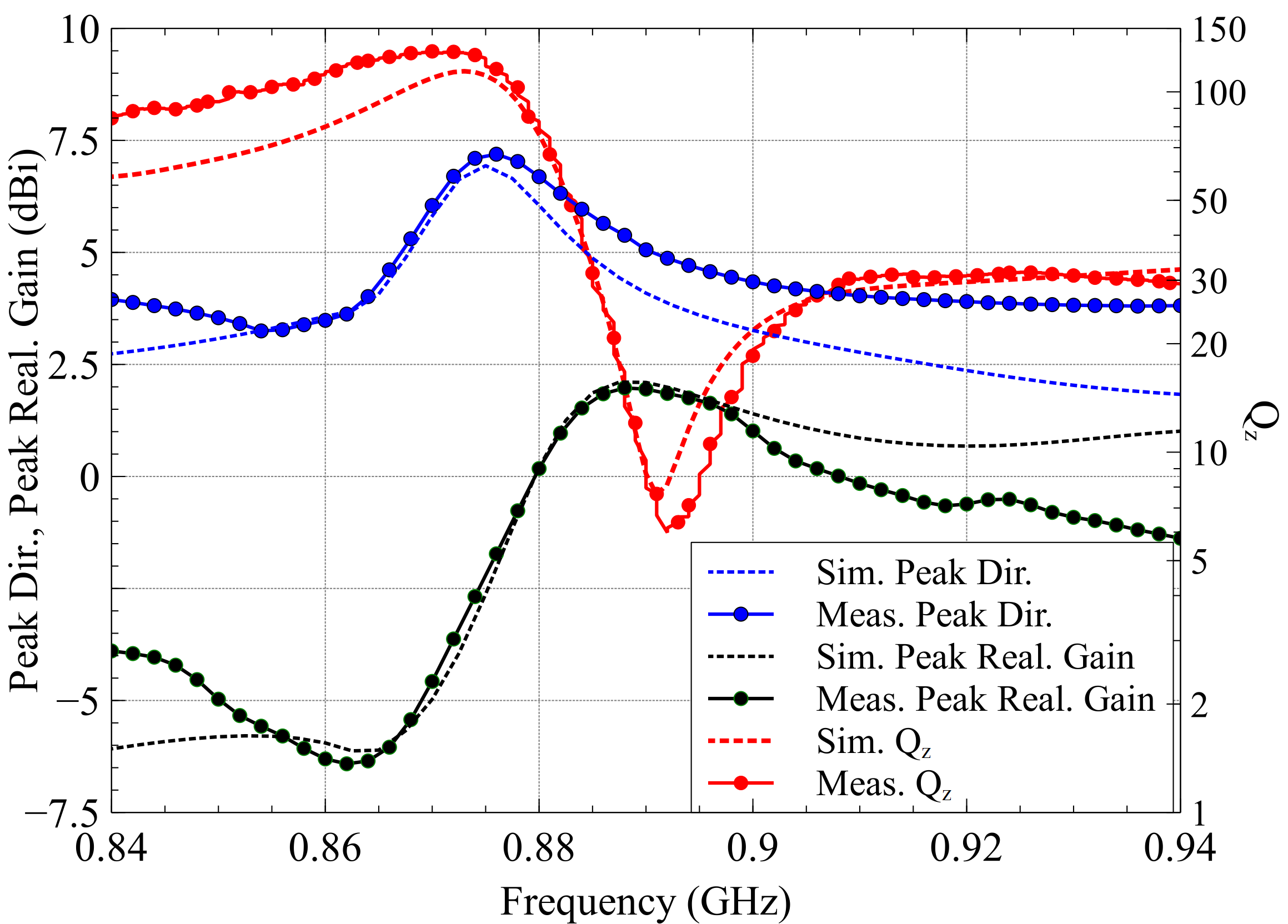}}
\caption{Measured and Simulated Peak directivity and Peak Gain.}
\label{Dir}
\end{figure}

\begin{figure}[t]
\centering{\includegraphics[width=85mm]{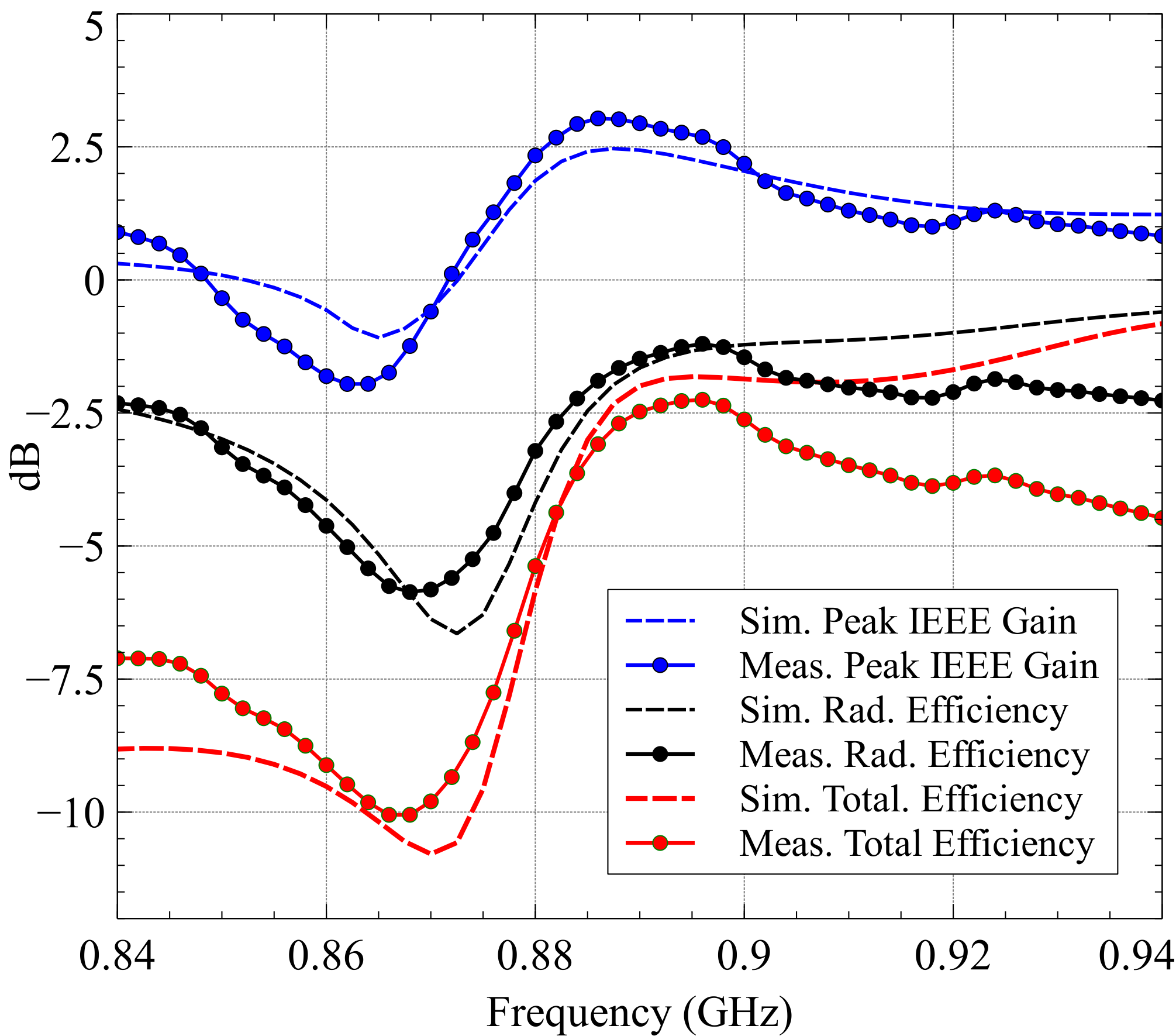}}
\caption{Measured and Simulated Peak IEEE Gain and radiated efficiency.}
\label{Eff}
\end{figure}

\begin{figure}[t]
\centering{
\subfigure[]{\includegraphics[width=25mm]{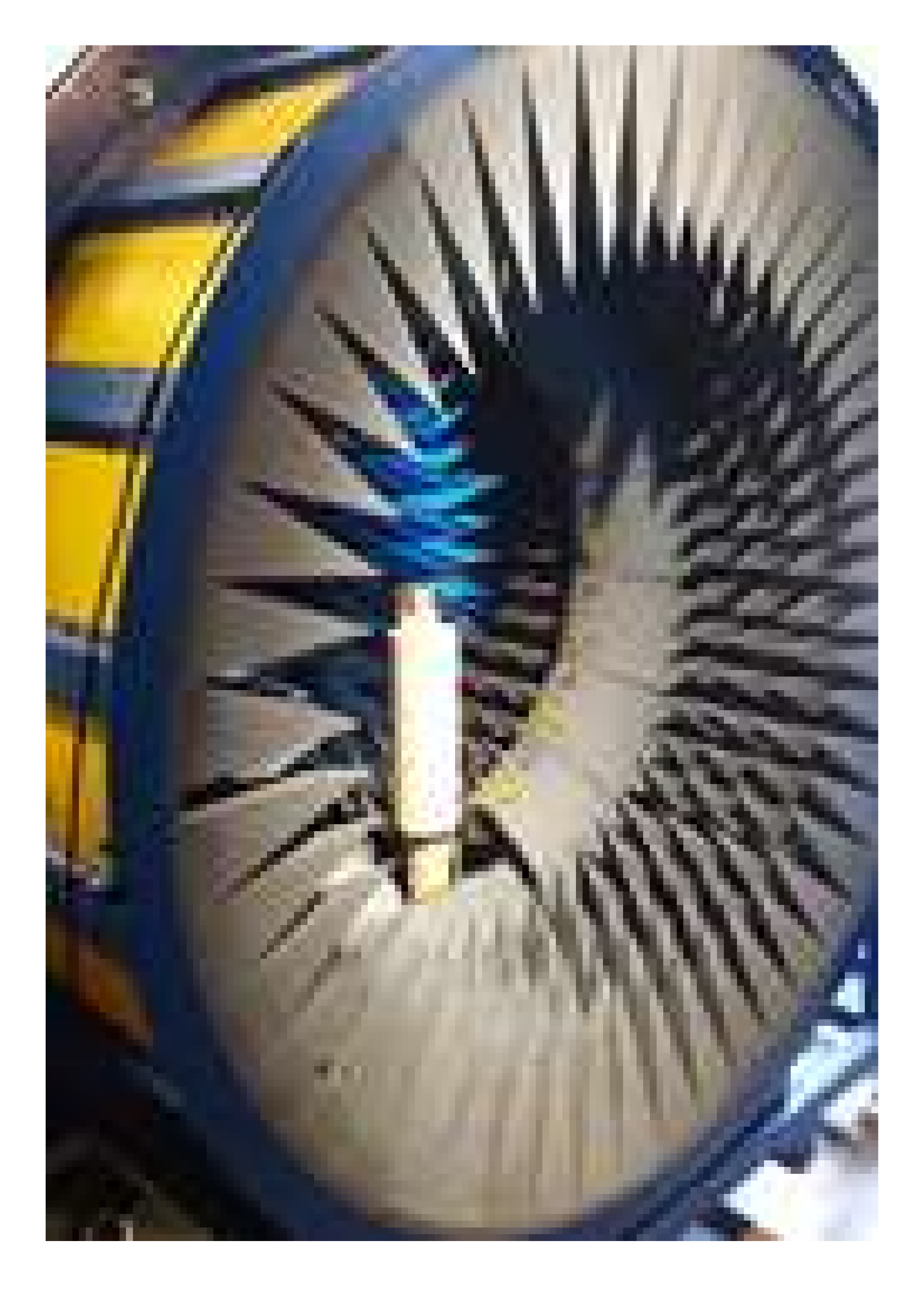}
\label{satimo}}
\subfigure[]{\includegraphics[width=55mm]{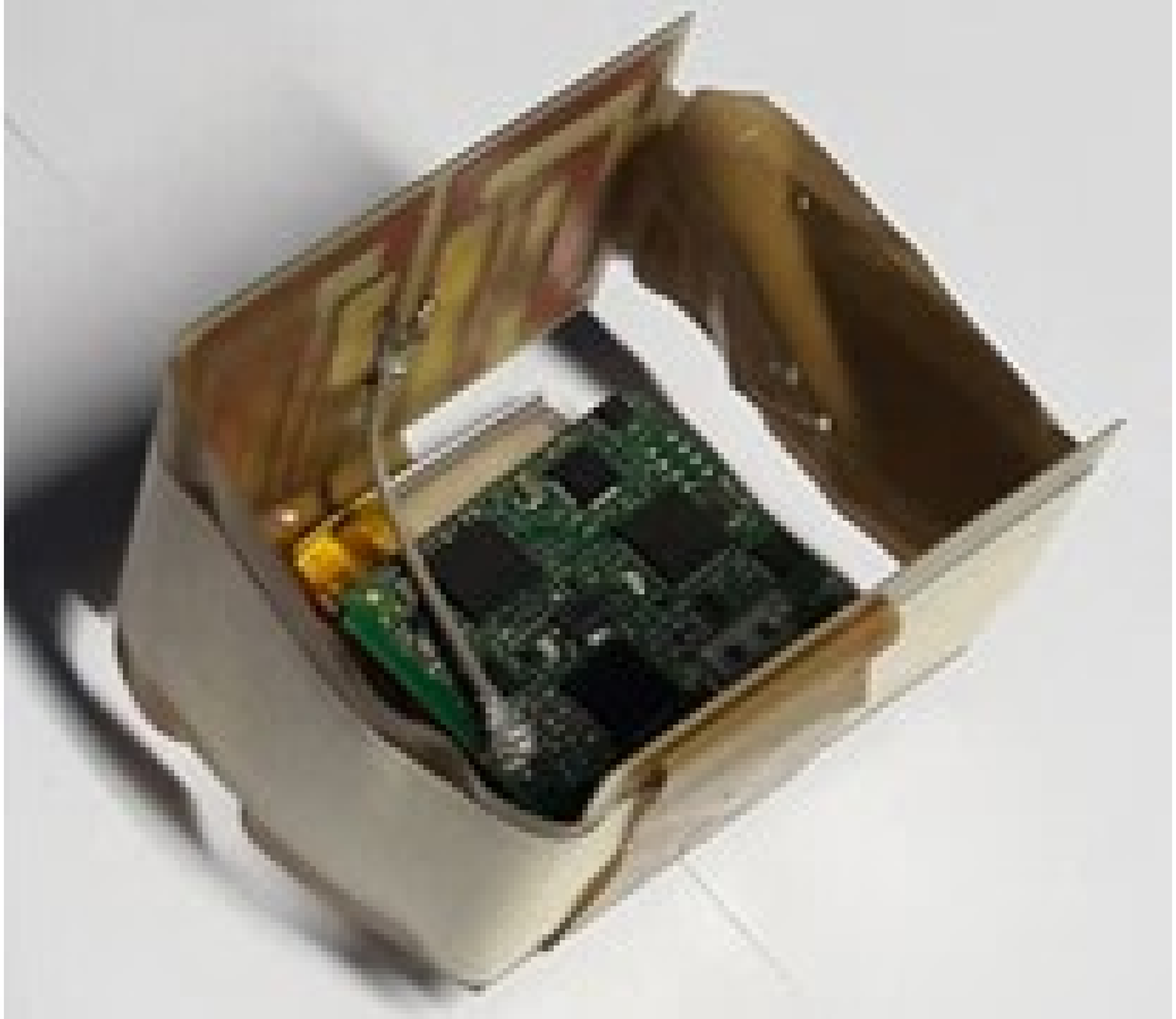}
\label{proto}}
\caption{ (a) Satimo station. (b) Autonomous radiating system.}\label{diag_mesure_sim}}
\end{figure}

The prototype radiation characteristics are measured on a SATIMO Starlab station and shown in Fig.~\ref{satimo}.
Peak directivity and realized gain are shown in Fig.~\ref{Dir}. The expected correlation between Directivity and $Q_Z$ is experimentally verified with this prototype.
Radiation and total efficiency, and Peak IEEE gain  are presented in Fig.~\ref{Eff}.
A peak directivity of 7 dBi is simulated at 0.875 GHz with a radiation efficiency of -6 dB. 
A good agreement is found with simulation considering a 120mm cable length. The measured peak directivity is 7.2 dBi with a -6.5 dB radiation efficiency at 0.876 GHz. The peak gain is obtained at 0.884 GHz. A very good agreement is found between the simulated and measured realized gain. However, some discrepancies appears for the directivity, IEEE gain and radiation efficiency. The measured directivity and radiation efficiency are very close to the simulated one.

\begin{figure}[t]
\centering{\includegraphics[width=85mm]{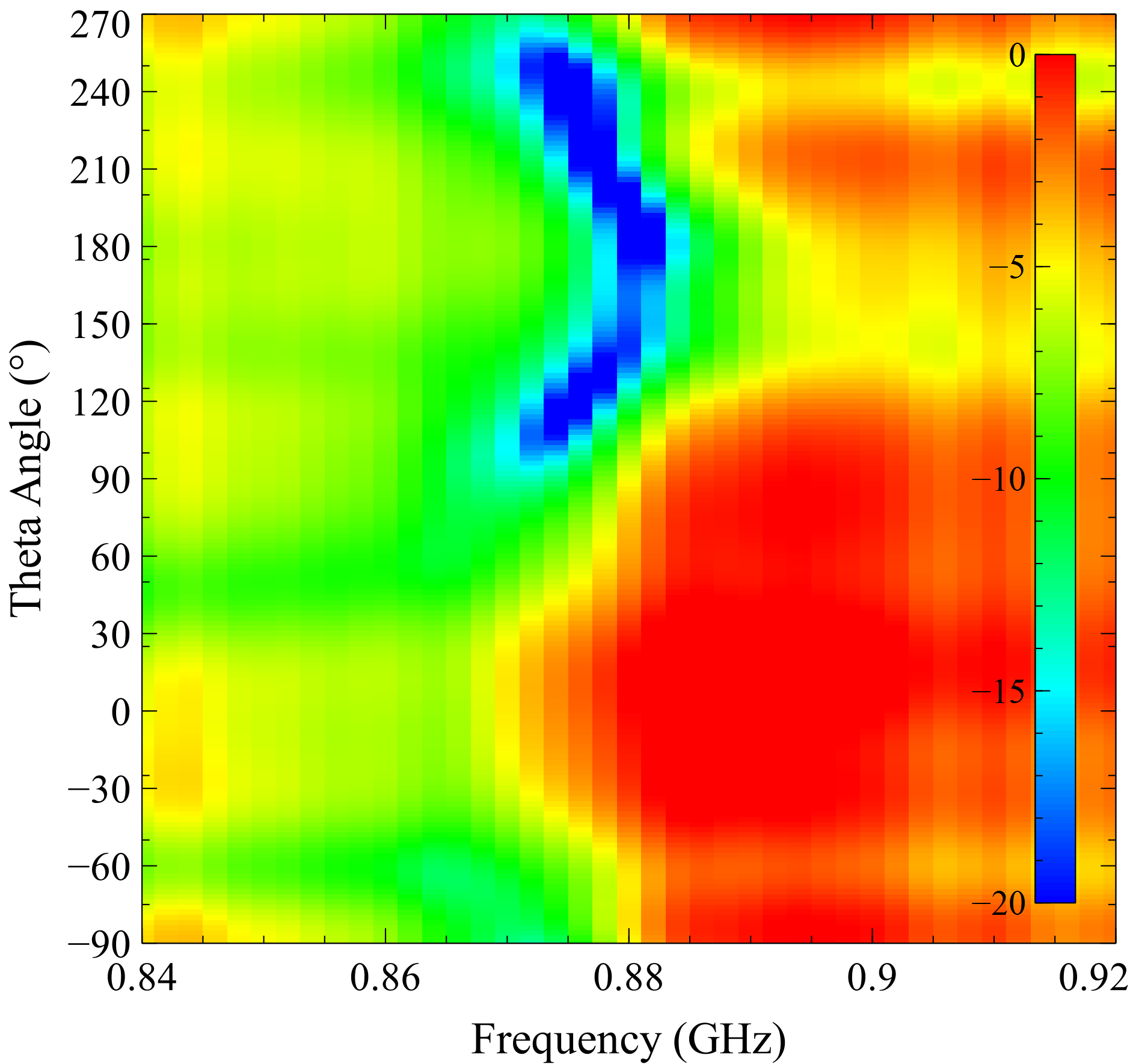}}
\caption{Measured total radiation gain versus frequency in the xOz plane.}
\label{xoz}
\end{figure}

\begin{figure}[t]
\centering{\includegraphics[width=85mm]{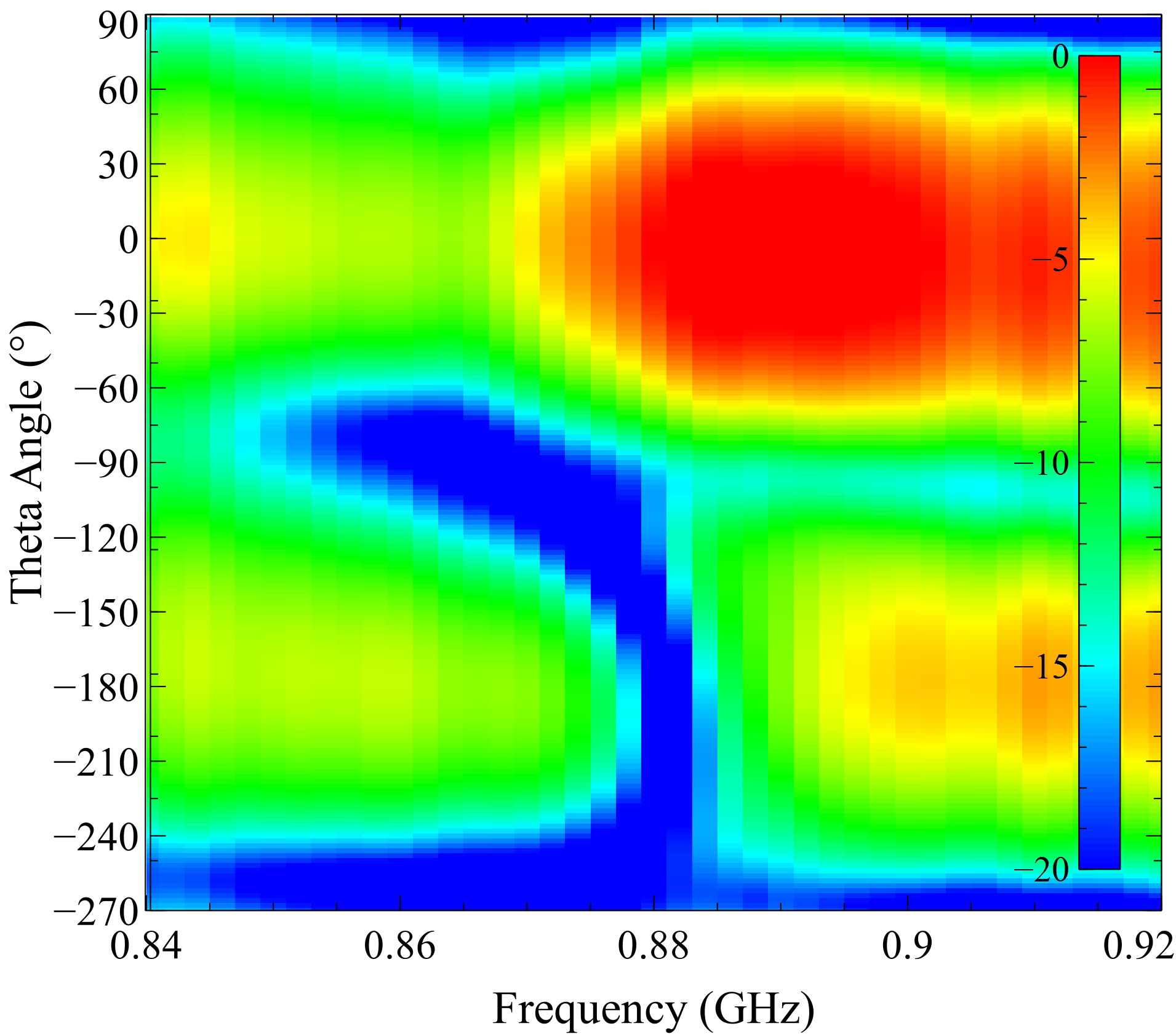}}
\caption{Measured total radiation gain versus frequency in the yOz plane. }
\label{yoz}
\end{figure}

In order to study the spatial filtering selectivity versus frequency, the total realized gain in the xOz and xOy plane are plotted versus frequency on Fig. \ref{xoz} and Fig. \ref{yoz}.  The optimal trade-off between peak realized gain and front-to-back ratio is obtained at 0.885 GHz (5.5dBi peak directivity). It can be seen that the spatial filtering mode is present for a very narrow frequency band.

\subsection{Measurement with an autonomous transciever}
As presented in Sec.~III, the $S_{11}$-effect of the cable is not negligible despite the balun correction effect. In order to assess the antenna performance without any cable effect and to demonstrate the structure capability to operate in a real communication application, a measurement with an autonomous emitter in continuous wave mode is performed. 
The PCB has a size of 32*25*1 mm and is powered by a LiPO Battery with a size of 25*25*5 mm. The electronic board is based on a SX1276 LoRa transciever from Semtech with integrated power amplifier. A UFL connector is used to connect the emitter to the antenna input. Before performing the measurement, the autonomous emitter was tested by connecting it to a power meter. A stable 13.7 dBm continuous wave is measured.

\begin{figure}[t]
\centering{\includegraphics[width=85mm]{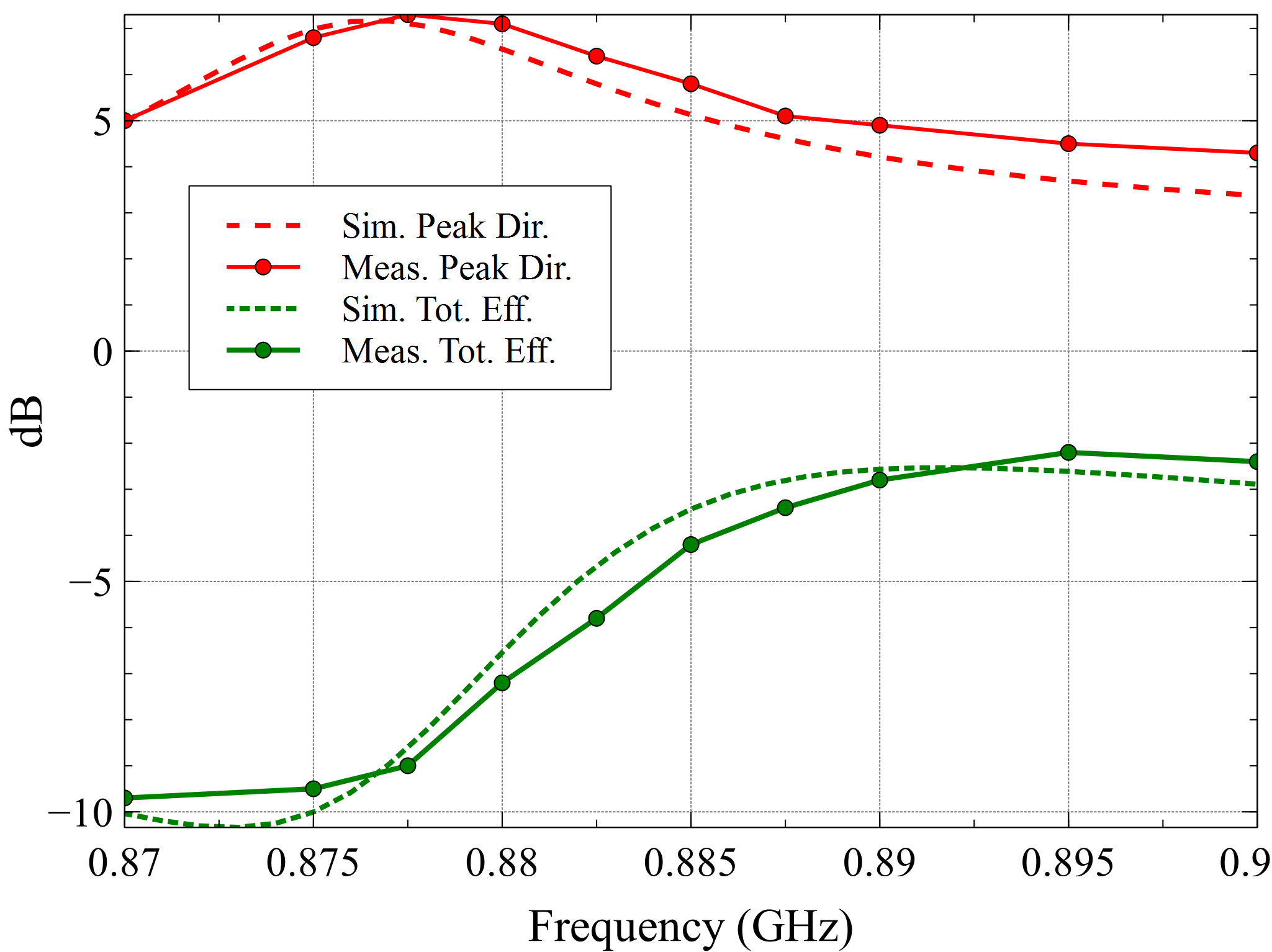}}
\caption{Measured Peak directivity and Total Efficiency versus frequency}
\label{Dir_eff}
\end{figure}

\begin{figure}[t]
\centering{\includegraphics[width=85mm]{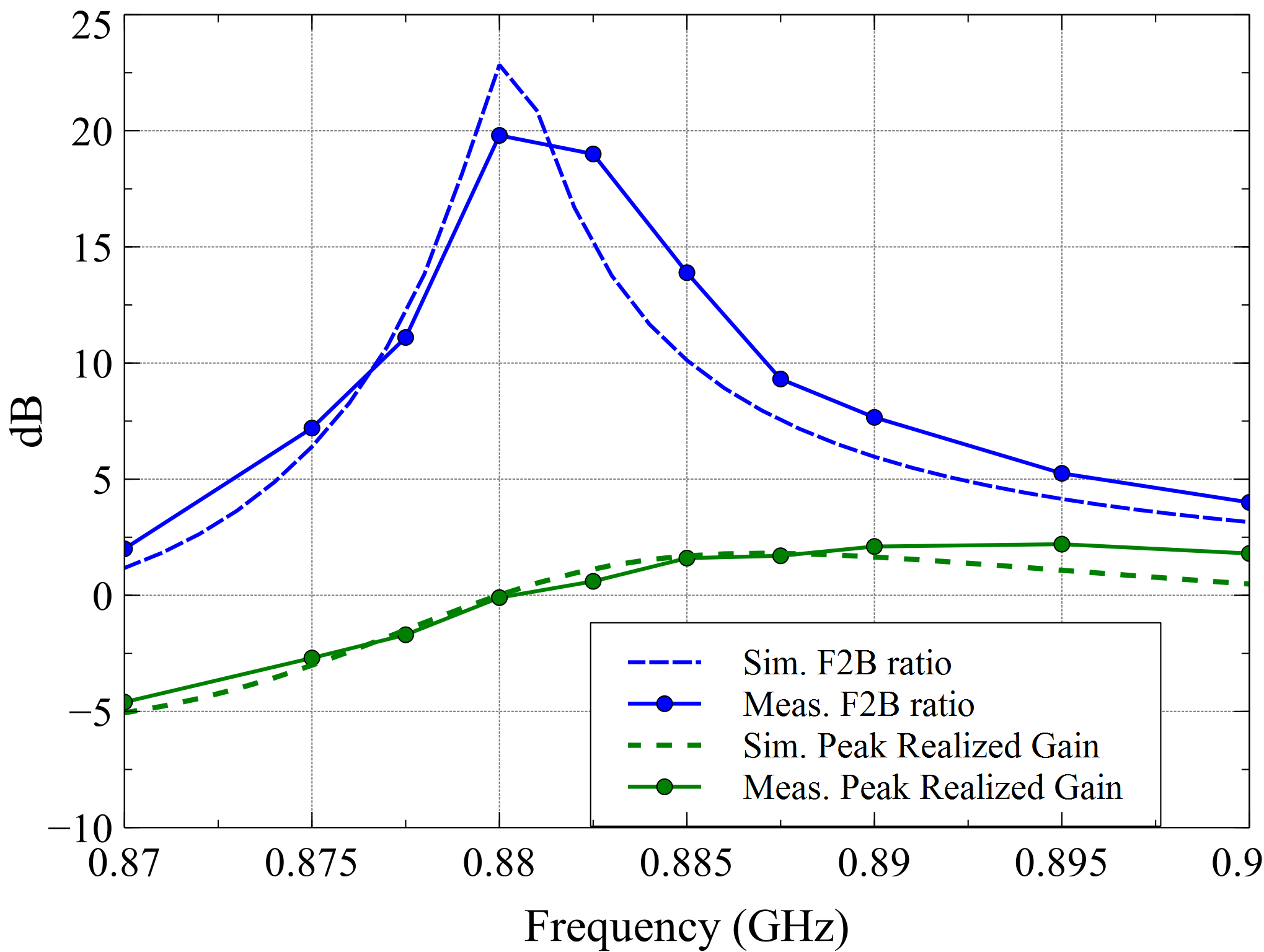}}
\caption{Measured Peak Realized Gain. and Front to back ratio versus frequency}
\label{Gain_F2B}
\end{figure}

The 3D Transmitted Radiated Power (TRP) is realized for 10 different frequencies. Between the different frequency measurement, the code of the emitter is updated with a micro-USB cable to change the center frequency.
Measurement of the peak directivity and total efficiency are presented on Fig. \ref{Dir_eff}. Total efficiency and realized gain are extracted using substitution method with a reference dipole antenna.
The measured realized gain and front-to-back ratio are presented on Fig. \ref{Gain_F2B}. 
A simulation of the antenna integrating the  emitter is performed and compared with measurement. A very good agreement is observed between simulation and measurement. As expected, the peak directivity and total efficiency are inversely proportional, and a trade-off between theses two parameters is required, see Fig.\ref{Dir_eff}. At 0.885 GHz, a good compromise between a 5.5 dBi peak directivity and -4 dB total efficiency is obtained.
A front to back ratio higher than 15 dB is observed from 0.878 to 0.884 GHz in Fig.\ref{Gain_F2B}.

\begin{figure}[t]
\centering{\includegraphics[width=85mm]{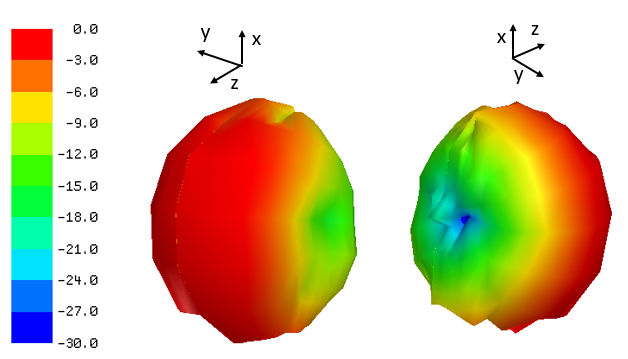}}
\caption{Measured 3D Total radiation pattern @880MHz}
\label{880_3D}
\end{figure}

The radiation pattern is investigated at 0.88 GHz because it is the frequency with the best spatial filtering capability (higher F2B ratio).
The 3D measured radiation pattern is presented on Fig. \ref{880_3D}.
From the 3D measurements, a 10.7 dB power ratio is computed between the hemisphere with $z>0$ and the hemisphere with $z<0$. Note that the spatial filtering is rather effective at 0.88 GHz.

\begin{figure}[t]
\centering{\includegraphics[width=85mm]{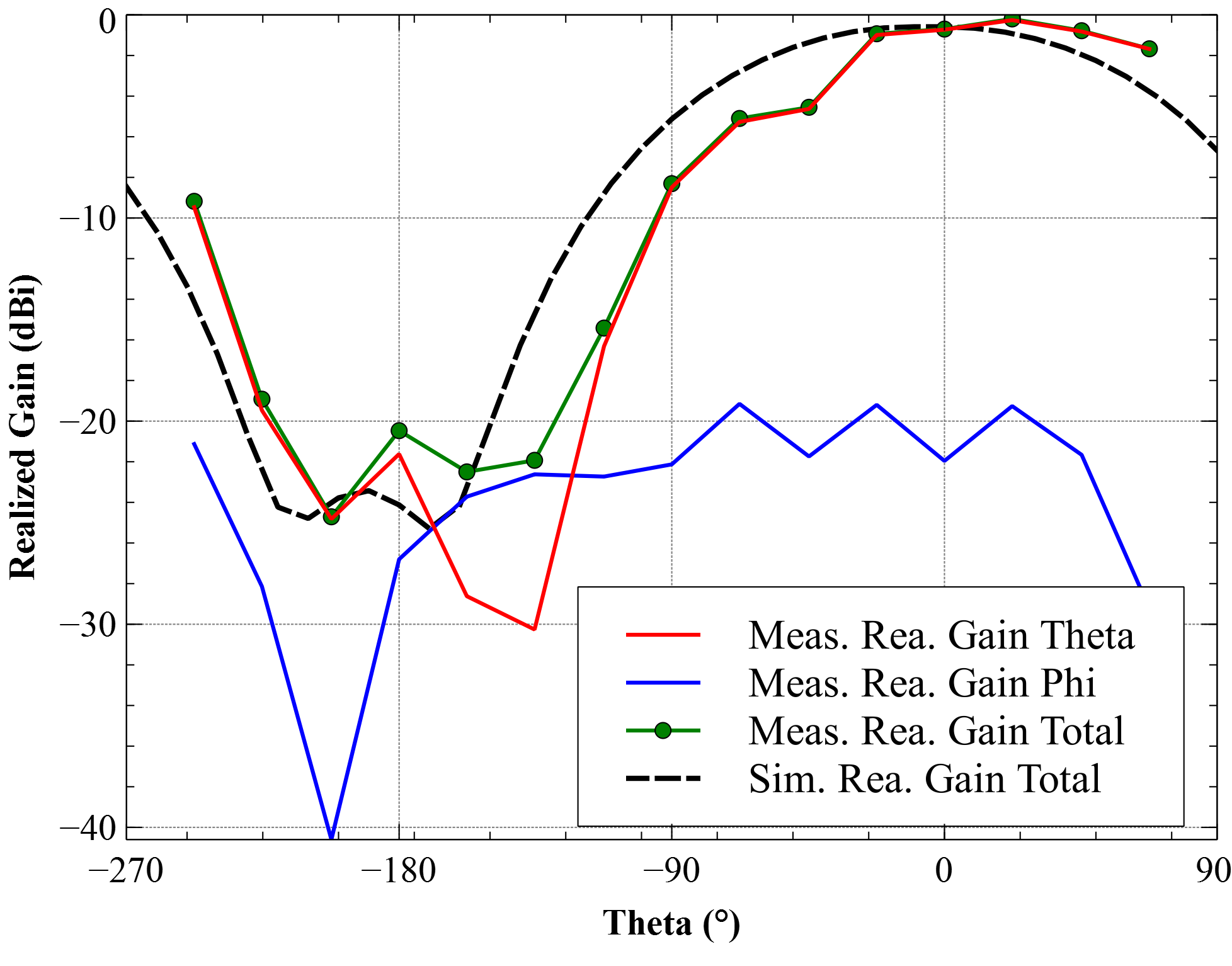}}
\caption{Measured Co, Cross and Total radiation pattern versus theta for $\phi=0^\circ$ @0.88 GHz (xOz plane)}
\label{880_xoz}
\end{figure}

\begin{figure}[t]
\centering{\includegraphics[width=85mm]{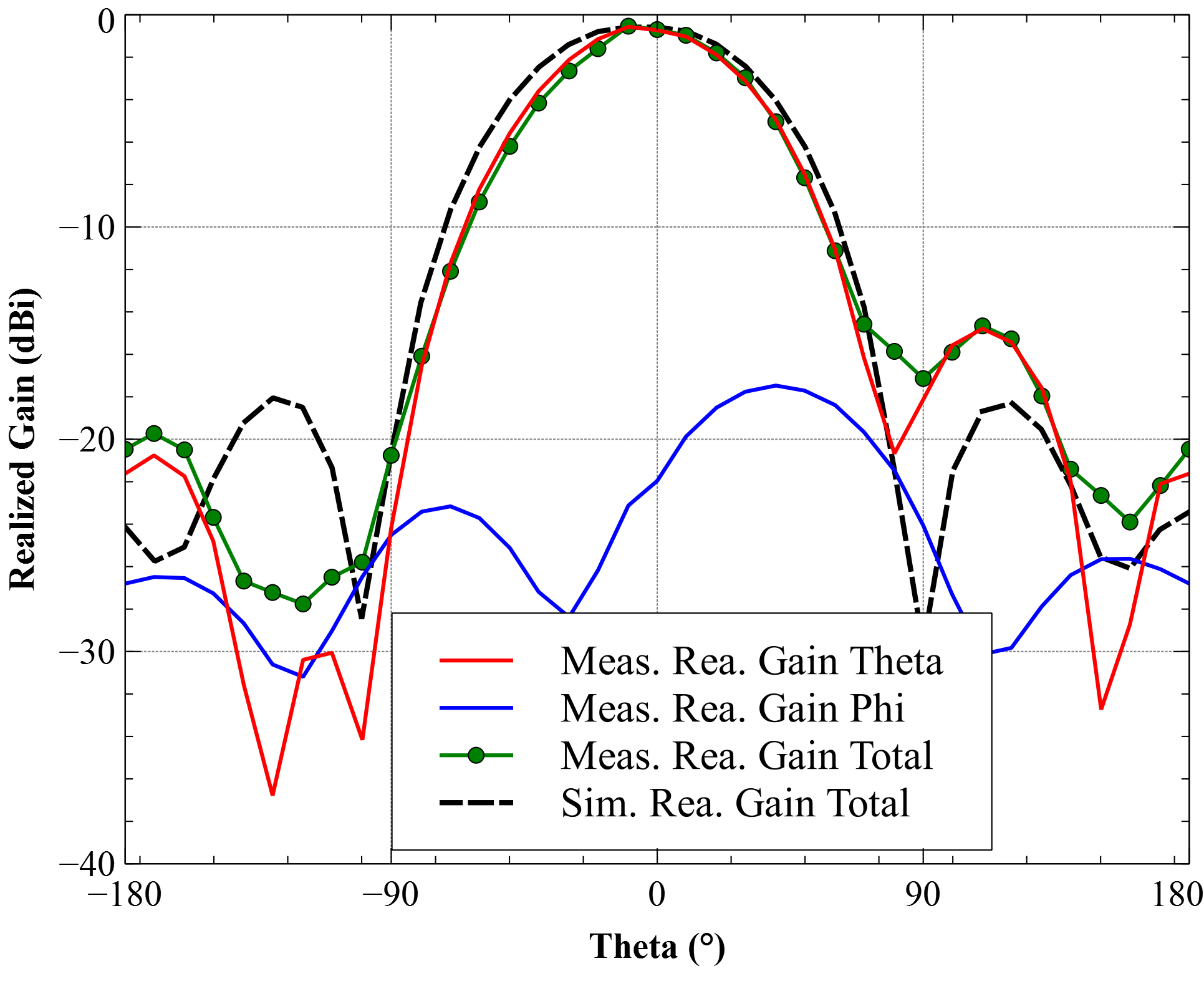}}
\caption{Measured Co, Cross and Total radiation pattern versus theta for $\phi=90^\circ$ @ 0.88 GHz(yOz plane)}
\label{880_yoz}
\end{figure}

The realized gain on xOz and yOz plane are plotted on Fig. \ref{880_xoz} and \ref{880_yoz}. On Fig. \ref{880_xoz}, the measured main beam is tilted to $+20^\circ$ in xOz plane when compared with simulated radiation pattern. This effect might be due to the influence of the PCB.
A fair agreement is found with simulation. The cross-polarization value is small which is expected for a miniature antenna.

\begin{figure}[t]
\centering{\includegraphics[width=85mm]{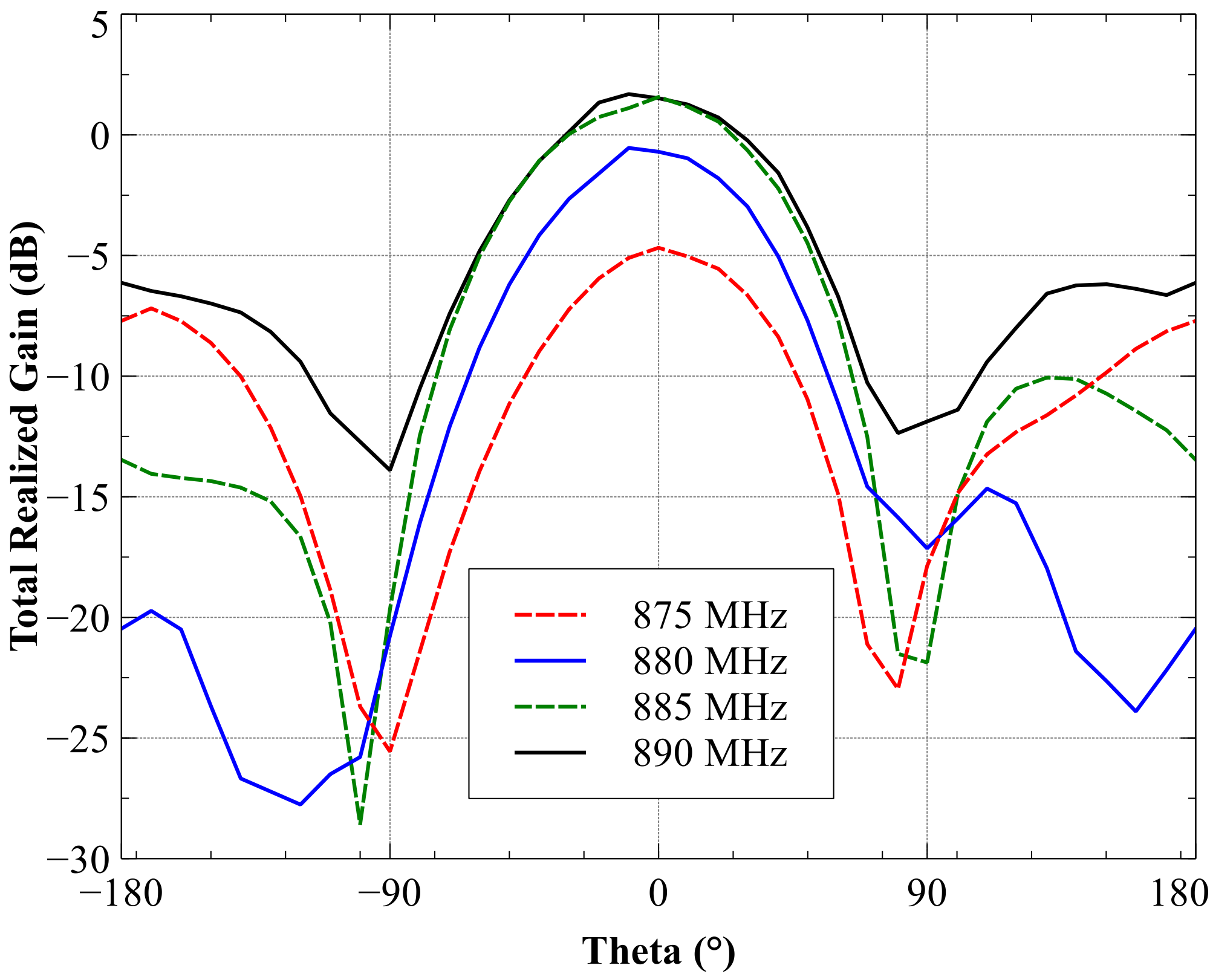}}
\caption{Measured Total radiation pattern versus theta for $\phi=90^\circ$ and 4 different frequencies (YoZ plane)}
\label{freq_yoz}
\end{figure}

In order to investigate the filtering capability versus frequency, the total radiation gain in yOz plane is depicted in Fig.~\ref{freq_yoz} for four different frequencies. The maximal gain is obtained for 0.89 GHz but an important back radiation lobe is observed. The front-to-back ratio is higher than 15dB for 0.88 and 0.885 GHz, implying that the a 5MHz bandwidth can be used. At 0.875 GHz, the gain is strongly degraded and the back radiation lobe is very high.

\section{Discussion}

In this paper we use a case study to show a practical approach to the optimization and design of a high directivity antenna, under a number of different constraints. 
The effect of conductive losses results in expected observations as the efficiency drop, and less obvious results like the front to back ratio improvement.
The comparison with fundamental limits show that the antenna is very close to the bounds for a moderate directivity (4 dBi). For higher directivity ($>$4.7dBi), Fig. ~\ref{Qz_supD} demonstrates that a more optimal shaping of the current could provide a one decade improvement on the quality factor.
The measurement, with and without the cable, shows a very good agreement between the two methods near antenna resonance (0.88-0.9 GHz). Some discrepancies appears outside this band due to balun amplitude and phase imbalance.

\section{Conclusion}

In this study, a super-directive parasitic element antenna structure is simulated and measured. Considering that the minimum sphere circumscribing the antenna is equal to $0.2 \lambda_0$, the proposed structure has a $ka < 0.7$.
The classical super-directive mode is observed with a large drop of the radiating efficiency at 0.876 GHz. However, the radiating efficiency drop can be mitigated using a slightly higher frequency. At 0.885 GHz, a -4dB radiating efficiency is combined with a 5.5 dBi directive radiation pattern  suitable for spatial filtering applications with a $Q_Z=35$. 

We show that we can utilize the stored energy bounds to predict a lower bound of the Q-factor as a function of total-directivity. The bounds indicates where the structure starts to become superdirective. The prediction for the given structure is that above $D\sim 4.5$ dBi, we see that the minimum bound on the Q-factor begins to grow rapidly. This is consistent with the measured observations. 

We also show that the proposed antenna structure and its properties are robust are compatible with the use of a PCB board and a battery, in such a way that superdirective properties are preserved when the antenna is in an autonomous setting. Thus the design has a robustness in its properties.

\section*{Acknowledgment}

The authors would like to thank the CREMANT for its support in measurements. The labex UCN@Sophia is also acknowledge for having funded Lars Jonsson visiting period in University Nice Sophia. FF would like to thank ERASMUS+ funding for visiting period in KTH Stockholm. LJ is grateful for support from SSF/AM130011 and LJ \& LW form Vinnova/ChaseOn in the project iAA.

\ifCLASSOPTIONcaptionsoff
  \newpage
\fi

\end{document}